\documentclass[11pt,english]{article}
\usepackage[T1]{fontenc}
\usepackage[latin9]{inputenc}
\usepackage{xcolor}
\usepackage{booktabs}
\usepackage{amsmath}
\usepackage{graphicx}
\usepackage[numbers]{natbib}

\makeatletter

\providecommand{\tabularnewline}{\\}

\usepackage[T1]{fontenc}
\usepackage[latin9]{inputenc}
\usepackage{color}
\usepackage{float}
\usepackage{graphicx}
\usepackage{graphics}
\usepackage{esint}
\usepackage{enumerate}

\makeatletter

\providecommand{\tabularnewline}{\\}

\usepackage{latexsym}
\usepackage{epsfig}
\usepackage{graphicx}
\usepackage[colorlinks,linkcolor=blue]{hyperref}
\allowdisplaybreaks 

\usepackage{babel}
\usepackage{tcolorbox}

\usepackage{tikz}

\makeatother

\usepackage{babel}
\begin{document}
{}~ \hfill\vbox{\hbox{CTP-SCU/2026008}}\break
\vskip 2.5cm
\centerline{\Large \bf T-duality of string charge density/bit threads correspondence} 
\vspace*{1.0ex}

\centerline{\Large \bf in type II supergravity} 

\vspace*{11.0ex}
\centerline{\large  Houwen Wu and Shuxuan Ying}
\vspace*{7.0ex}
\vspace*{3.0ex}

\centerline{\large \it College of Physics}
\centerline{\large \it Sichuan University}
\centerline{\large \it Chengdu, 610065, China} \vspace*{1.0ex}

\vspace*{3.0ex}
\centerline{\large \it Department of Physics}
\centerline{\large \it Chongqing University}
\centerline{\large \it Chongqing, 401331, China} \vspace*{1.0ex}
\vspace*{3.0ex}

\centerline{iverwu@scu.edu.cn, ysxuan@cqu.edu.cn}
\vspace*{8.0ex}
\centerline{\bf Abstract} \bigskip \smallskip
In this paper, we study Abelian T-duality of the string charge density/bit threads correspondence in ten-dimensional Type II supergravity. T-duality along the common compact circle maps the Type IIB F1-NS5-P background to a Type IIA P-NS5-F1 background with the winding and momentum charges exchanged. The radial F1 source remains an F1 source, and its projected charge density gives a bit-thread flow whose maximal flux reproduces the black hole entropy. The same duality maps D1-D5-P to D0-D4-F1. In this case, a radial D1 source becomes a D2 source wrapped on the dual circle. Contracting the resulting spatial bivector with the normalized closed one-form on that circle gives a conserved effective radial flow whose maximal flux again reproduces the entropy. Although the local metrics, dilatons, source currents, and flow norms change, the dilaton-weighted transverse area density and the integrated maximal flux are invariant. These results provide a nontrivial T-duality test of the correspondence and extend it from string currents to wrapped-brane currents that reduce to string-like flows.

\vfill 
\eject
\baselineskip=16pt
\vspace*{10.0ex}
\tableofcontents

\section{Introduction}

\label{sec:introduction}

The Ryu--Takayanagi prescription relates the entropy of a region
in a holographic conformal field theory to the area of a bulk extremal
surface \cite{Ryu:2006bv,Ryu:2006ef}. On a static bulk slice, the
same entropy can be written as the maximal flux of a divergenceless
vector field \cite{Freedman:2016zud,Headrick:2017ucz}, 
\begin{equation}
\nabla_{i}v^{i}=0,\qquad|v|\leq\frac{1}{4G_{N}}.\label{eq:standard-bit-thread-bound}
\end{equation}
This formulation treats quantum information as a conserved flow. For
further developments and applications of the bit-thread formalism,
see Refs. \cite{Agon:2018lwq,Caggioli:2024uza,Lin:2026ehl,Caceres:2025ypk}.
However, the vector field is normally introduced as an auxiliary geometric
object, and its microscopic carrier is not fixed by the max-flow/min-cut
theorem.

In Ref. \cite{Wu:2025qwc}, a conserved string charge density was
proposed as a physical realization of a distinguished bit-thread flow.
The antisymmetric spacetime current obtained by varying the worldsheet
coupling to the Kalb--Ramond field has components $j^{0\mathrm{a}}$
tangent to the string. Their projection onto a constant-time slice
is divergenceless and can be mapped to a bit-thread vector. This construction
reproduces the entropy of the BTZ black hole from a worldsheet-derived
observable. A subsequent analysis extended the correspondence to the
ten-dimensional F1--NS5--P and D1--D5--P systems and tested it
under Type IIB S-duality \cite{Wu:2026qha}. S-duality changes an
F1 carrier into a D1 carrier while preserving the rank of the two-form
current.

T-duality gives a different and more demanding test. It mixes the
metric and the Kalb--Ramond field, changes the compactification radius
and the string coupling, and can change the dimension of a D-brane.
In particular, a T-duality transverse to a radial D1 probe maps it
to a D2 probe wrapped on the dual circle \cite{Bergshoeff:1996tu,Hassan:1999bv}.
Therefore, a T-duality covariant formulation cannot assume that the
carrier is always described by a two-form spacetime current. It must
explain how a higher-rank brane current can still define the one-dimensional
flow required by bit threads.

In this paper, we study T-duality along the common compact direction
$w$ of the three-charge systems. We consider two duality chains:
\begin{align}
\mathrm{IIB}:\quad\mathrm{F1}\left(w\right)-\mathrm{NS5}\left(wT^{4}\right)-\mathrm{P}\left(w\right) & \xrightarrow{\ T_{w}\ }\mathrm{IIA}:\quad\mathrm{P}\left(\widetilde{w}\right)-\mathrm{NS5}\left(\widetilde{w}T^{4}\right)-\mathrm{F1}\left(\widetilde{w}\right),\label{eq:NS-chain-intro}\\
\mathrm{IIB}:\quad\mathrm{D1}\left(w\right)-\mathrm{D5}\left(wT^{4}\right)-\mathrm{P}\left(w\right) & \xrightarrow{\ T_{w}\ }\mathrm{IIA}:\quad\mathrm{D0}-\mathrm{D4}\left(T^{4}\right)-\mathrm{F1}\left(\widetilde{w}\right).\label{eq:RR-chain-intro}
\end{align}
In the first line, the integer winding and momentum charges are exchanged.
In the second line, the background D1 and D5 charges become D0 and
D4 charges, while the momentum becomes F1 winding.

We first apply the Buscher rules to the complete stationary backgrounds,
including the off-diagonal component $g_{tw}$ and the electric two-form
component $B_{tw}$. For the NS--NS chain, the T-dual background
has the same functional form as P--NS5--F1, with the harmonic functions
of F1 winding and momentum exchanged. A smeared radial F1 current
gives a conserved flow, and its maximal flux reproduces the Bekenstein--Hawking
entropy. The related studies have investigated black hole entropy
from the perspective of string worldsheet theory \cite{He:2014gva,Ahmadain:2022tew,Ahmadain:2022eso,Brustein:2022wiq,Halder:2023adw,Ahmadain:2024hdp,Mori:2025qoh,Ahmadain:2025pox,Jorstad:2026jlg}.

\begin{figure}[h]
\begin{centering}
\includegraphics[scale=0.22]{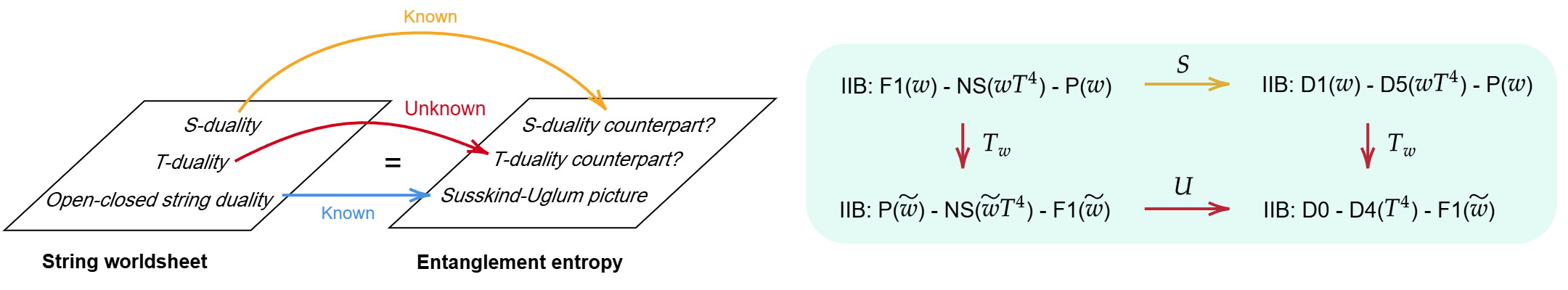}
\par\end{centering}
\caption{\label{fig:aim}The \textbf{\textcolor{blue}{blue path}} represents
our original three-dimensional construction of the string charge density/bit
threads correspondence, in which entanglement entropy is computed
from an open-string charge density, while the Bekenstein--Hawking
entropy is encoded by the corresponding closed-string winding charge
through open--closed string duality \cite{Wu:2025qwc}. The \textbf{\textcolor{orange}{orange
path}} represents our extension of this correspondence to ten-dimensional
Type IIB supergravity and its consistency under S-duality \cite{Wu:2026qha}.
The \textbf{\textcolor{red}{red path}}, which is the focus of the
present paper, illustrates how the correspondence transforms under
T-duality. The map connecting the two Type IIA configurations on the
right-hand side is a U-duality transformation. It is not an additional
duality imposed after T-dualization; rather, it is precisely the representation
of the original type IIB S-duality in the T-dual frame, $U=T_{w}ST_{w}^{-1}$.}
\end{figure}

Then, we transform the D1--D5--P background and its radial D1 source.
The background becomes D0--D4--F1, while each radial D1 probe becomes
a D2 probe wrapped on $S_{\widetilde{w}}^{1}$. The spatial projection
of the D2 current is a bivector. Let 
\begin{equation}
\eta\equiv\frac{d\widetilde{w}}{L_{\widetilde{w}}},\qquad\int_{S_{\widetilde{w}}^{1}}\eta=1,\label{eq:normalized-eta-intro}
\end{equation}
be the normalized one-form on the dual circle. Contracting the projected
D2 current with $\eta$ gives a conserved radial vector. The resulting
vector has total charge flux $\mathcal{N}$, independent of the radius
of the dual circle. The logic of this paper is therefore summarized
in Fig. (\ref{fig:aim}).

Finally, we identify the local quantity that controls the entropy
flow in every duality frame. If $Y(r)$ is the string-frame area density
of a constant-$r$ horizon section, with the radial metric factor
removed, then 
\begin{equation}
\mathcal{A}\left(r\right)\equiv e^{-2\phi\left(r\right)}Y\left(r\right)=r^{3}\sqrt{H_{1}\left(r\right)H_{5}\left(r\right)H_{P}\left(r\right)},\label{eq:universal-weighted-area-intro}
\end{equation}
in all four descriptions. The Buscher transformation changes $Y$
and the dilaton separately but leaves their product unchanged. Globally,
the transformation of the dual radius and the ten-dimensional Newton
constant gives 
\begin{equation}
\frac{\widetilde{\mathcal{V}}_{0}}{\widetilde{G}_{N}^{\left(10\right)}}=\frac{\mathcal{V}_{0}}{G_{N}^{\left(10\right)}}.\label{eq:global-invariant-intro}
\end{equation}
These two identities imply the invariance of the normalized flow and
of the integrated maximal entropy flux.

The calculation of this paper is performed at two-derivative order
and in the probe and continuum-smearing approximations. The source
ensembles are independent of the dualized coordinate, as required
for the supergravity Buscher rules. A fully localized T-dual source
would require a string-scale description beyond the zero-mode truncation.
Subject to these assumptions, the result provides a direct test that
the string charge density/bit threads correspondence is compatible
with leading-order Abelian T-duality.

The paper is organized as follows. Section \ref{sec:setup} reviews
the projected string charge density, the string-frame norm bound,
and the Buscher rules. Section \ref{sec:NS-dual} studies the NS--NS
duality chain and its radial F1 flow. Section \ref{sec:RR-dual} studies
D1--D5--P, its D0--D4--F1 dual, and the wrapped-D2 current. Section
\ref{sec:invariants} compares the local and global flows in all frames.
Section \ref{sec:conclusion} gives the physical interpretation and
discusses limitations. 

\section{String charge flows and Abelian T-duality}

\label{sec:setup}

In this section, we first briefly review the correspondence between
string charge density and bit-thread flow and explain how black hole
entropy, or equivalently the maximal entropy flux, can be obtained
from the string charge density. We then review the Type IIB F1--NS5--P
and D1--D5--P backgrounds, together with the Buscher rules.

\subsection{Projected string charge density and bit-thread flow}

\label{subsec:projected-current}

The fundamental string source that couples to the Kalb--Ramond field
is
\begin{equation}
S_{\mathrm{F1}}=-\frac{1}{4\pi\alpha^{\prime}}\int d^{2}\sigma\left[\sqrt{-\gamma}\gamma^{mn}\partial_{m}X^{\mathrm{a}}\partial_{n}X^{\mathrm{b}}g_{\mathrm{ab}}+\epsilon^{mn}\partial_{m}X^{\mathrm{a}}\partial_{n}X^{\mathrm{b}}B_{\mathrm{ab}}\right].\label{eq:F1-Polyakov-action}
\end{equation}
Varying the second term with respect to $B_{\mathrm{ab}}$ gives the
antisymmetric spacetime current 
\begin{equation}
j^{\mathrm{ab}}\left(x\right)=\frac{1}{\sqrt{-g}}\int d^{2}\sigma\left(\partial_{\tau}X^{\mathrm{a}}\partial_{\sigma}X^{\mathrm{b}}-\partial_{\tau}X^{\mathrm{b}}\partial_{\sigma}X^{\mathrm{a}}\right)\delta^{\left(10\right)}\left(x-X\left(\tau,\sigma\right)\right).\label{eq:F1-current-general}
\end{equation}
It obeys 
\begin{equation}
\partial_{\mathrm{a}}\left(\sqrt{-g}j^{\mathrm{ab}}\right)=0,\qquad\mathrm{a}=x^{0},\ldots,x^{9},\label{eq:spacetime-current-conservation}
\end{equation}
away from endpoints. On a stationary constant-time slice with induced
metric $h_{IJ}$ and ADM lapse $N_{{\rm ADM}}$, define 
\begin{equation}
q^{I}\equiv N_{{\rm ADM}}j^{0I}=\frac{\sqrt{-g}}{\sqrt{h}}j^{0I},\qquad I=x^{1},\ldots,x^{9},\label{eq:projected-current}
\end{equation}
where we denote $j^{\mathrm{ab}}$ by the string current, $j^{0I}$
by the string charge density, and $q^{I}$ by the projected string
charge density. For a source whose projected string charge density
has no time dependence on the chosen slice, 
\begin{equation}
\partial_{I}\left(\sqrt{h}q^{I}\right)=0.\label{eq:projected-conservation}
\end{equation}
We associate to this projected string charge density the bit-thread
flow 
\begin{equation}
v^{I}=C_{{\rm geom}}q^{I}.\label{eq:flow-from-current}
\end{equation}
The coefficient $C_{{\rm geom}}$ converts normalized source charge
density into entropy flux.

The standard norm bound is imposed in the ten-dimensional Einstein
frame. We work with the normalized radial dilaton, setting the asymptotic
value to zero; the asymptotic string coupling is restored in $G_{N}^{\left(10\right)}$.
The Einstein metric normalized to agree with the string metric at
infinity is 
\begin{equation}
g_{\mathrm{ab}}^{{\rm E}}=e^{-\phi/2}g_{\mathrm{ab}}^{{\rm string}}.\label{eq:Einstein-string-frame}
\end{equation}
An eight-dimensional cut therefore obeys $dA_{{\rm E}}=e^{-2\phi}dA_{{\rm s}}$.
Requiring the entropy flux to be frame independent gives the string-frame
bound 
\begin{equation}
|v\left(r\right)|\leq\frac{e^{-2\phi\left(r\right)}}{4G_{N}^{\left(10\right)}}.\label{eq:string-frame-bound}
\end{equation}
This is also the local density that follows from the two-derivative
string-frame Wald entropy. Because the complete three-charge exteriors
are asymptotically flat rather than asymptotically AdS, we use bit-thread
flow below for a divergenceless entropy-flow vector obeying this local
capacity bound, with the horizon as the relevant cut. The standard
boundary-entanglement interpretation applies in the decoupled $\mathrm{AdS}_{3}$
throat.

For the homogeneous radial configurations used below, let the constant-$r$
section be 
\begin{equation}
\Sigma_{8}=S_{w}^{1}\times S^{3}\times T^{4},\qquad\mathcal{V}_{0}=L_{w}V_{T^{4}}\Omega_{3},\qquad\Omega_{3}=2\pi^{2}.\label{eq:reference-volume}
\end{equation}
The reference volume $\mathcal{V}_{0}$ is defined with the coordinate
measure and is independent of the dynamical metric. We write 
\begin{equation}
\sqrt{h}\equiv\sqrt{g_{rr}}\sqrt{g_{\Sigma_{8}}}=\sqrt{g_{rr}}Y\left(r\right)\sqrt{\bar{\gamma}},\label{eq:Y-definition}
\end{equation}
where $\bar{\gamma}$ is the determinant of the unit $S^{3}$ metric.
In the continuum-smearing approximation, a large uniform ensemble
of $\mathcal{N}$ radial probe sources, with total probe backreaction
parametrically smaller than that of the background, gives
\begin{equation}
q^{r}\left(r\right)=\frac{1}{\sqrt{g_{rr}}Y\left(r\right)}\frac{\mathcal{N}}{\mathcal{V}_{0}},\qquad\left|v\left(r\right)\right|=\frac{C_{{\rm geom}}}{Y\left(r\right)}\frac{\mathcal{N}}{\mathcal{V}_{0}}.\label{eq:universal-radial-current}
\end{equation}
Saturation of Eq. (\ref{eq:string-frame-bound}) at the extremal horizon
fixes 
\begin{equation}
C_{{\rm geom}}=\frac{\mathcal{V}_{0}}{4G_{N}^{\left(10\right)}\mathcal{N}}\lim_{r\to0^{+}}\left[e^{-2\phi\left(r\right)}Y\left(r\right)\right].\label{eq:universal-Cgeom}
\end{equation}
The flux through any constant-$r$ section is then 
\begin{equation}
\Phi\left(r\right)=\int_{\Sigma_{8}}v^{r}n_{r}dA=C_{{\rm geom}}\mathcal{N}.\label{eq:universal-flux}
\end{equation}

\subsection{The Type IIB three-charge backgrounds and Buscher rules}

\label{subsec:IIB-backgrounds}

To present the Type IIB three-charge backgrounds, we use a common
set of harmonic functions, 
\begin{equation}
H_{1}=1+\frac{Q_{1}}{r^{2}},\qquad H_{5}=1+\frac{Q_{5}}{r^{2}},\qquad H_{P}=1+\frac{Q_{P}}{r^{2}},\qquad K\equiv H_{P}-1.\label{eq:harmonic-functions}
\end{equation}
The Type IIB F1--NS5--P solution in the string frame is \cite{Tseytlin:1996bh,Mathur:2005zp}
\begin{align}
dS_{\mathrm{F1\text{\textendash}NS5\text{\textendash}P}}^{2} & =H_{1}^{-1}\left[-dt^{2}+dw^{2}+K\left(dt+dw\right)^{2}\right]+H_{5}\left(dr^{2}+r^{2}d\Omega_{3}^{2}\right)+dy_{i}dy^{i},\label{eq:IIB-F1-metric}\\
e^{2\phi_{\mathrm{F1}}} & =\frac{H_{5}}{H_{1}},\qquad B_{2}=\left(1-H_{1}^{-1}\right)dt\wedge dw+2Q_{5}\omega_{2},\qquad d\omega_{2}=\omega_{3}.\label{eq:IIB-F1-fields}
\end{align}
The F1 strings and NS5-branes share $w$, the NS5-branes also wrap
$T^{4}$, and the momentum wave propagates along $w$.

The S-dual D1--D5--P solution is 
\begin{align}
dS_{\mathrm{D1\text{\textendash}D5\text{\textendash}P}}^{2} & =\left(H_{1}H_{5}\right)^{-1/2}\left[-dt^{2}+dw^{2}+K\left(dt+dw\right)^{2}\right]\nonumber \\
 & \quad+\left(H_{1}H_{5}\right)^{1/2}\left(dr^{2}+r^{2}d\Omega_{3}^{2}\right)+\left(\frac{H_{1}}{H_{5}}\right)^{1/2}dy_{i}dy^{i},\label{eq:IIB-D1-metric}\\
e^{2\phi_{\mathrm{D1}}} & =\frac{H_{1}}{H_{5}},\label{eq:IIB-D1-dilaton}\\
C_{2} & =\left(1-H_{1}^{-1}\right)dt\wedge dw+2Q_{5}\omega_{2},\qquad B_{2}=0.\label{eq:IIB-D1-fields}
\end{align}
Both solutions have an extremal horizon at $r=0$. Compactification
on $S_{w}^{1}\times T^{4}$ gives a five-dimensional three-charge
black hole.

Assume that all background fields are independent of $w$. In local
coordinates in which the asymptotic metric component is $g_{ww}=1$,
the string-frame Buscher rules are \cite{Buscher:1987sk,Buscher:1987qj,Alvarez:1994dn}
\begin{align}
\widetilde{g}_{ww} & =\frac{1}{g_{ww}}, & \widetilde{g}_{\mu w} & =\frac{B_{\mu w}}{g_{ww}}, & \widetilde{B}_{\mu w} & =\frac{g_{\mu w}}{g_{ww}},\label{eq:Buscher-1}\\
\widetilde{g}_{\mu\nu} & =g_{\mu\nu}-\frac{g_{\mu w}g_{\nu w}-B_{\mu w}B_{\nu w}}{g_{ww}}, & \widetilde{B}_{\mu\nu} & =B_{\mu\nu}-\frac{B_{\mu w}g_{\nu w}-g_{\mu w}B_{\nu w}}{g_{ww}},\label{eq:Buscher-2}\\
e^{2\widetilde{\phi}} & =e^{2\phi}/g_{ww}.\label{eq:Buscher-dilaton}
\end{align}
Here $\mu,\nu$ denote all directions other than $w$. These equations
also fix the orientation convention for $\widetilde{w}$. There is
no additional reversal of the dual coordinate is made below. Convention
dependent signs in the R--R sector will be fixed separately by the
orientation of the corresponding D-brane charges.

Eq. (\ref{eq:Buscher-dilaton}) is written for the radial dilaton
normalized to vanish at infinity. The constant dilaton zero mode is
recorded separately by the transformation of the asymptotic coupling
below. This separation avoids mixing the local Buscher factor $g_{ww}\left(r\right)$
with the change of the coordinate radius.

Let $w\sim w+2\pi R_{w}$. Restoring $\alpha^{\prime}$, the asymptotic
moduli transform as 
\begin{equation}
R_{\widetilde{w}}=\frac{\alpha^{\prime}}{R_{w}},\qquad\widetilde{g}_{s}=g_{s}\frac{\sqrt{\alpha^{\prime}}}{R_{w}}.\label{eq:T-dual-moduli}
\end{equation}
Since 
\begin{equation}
G_{N}^{\left(10\right)}=8\pi^{6}g_{s}^{2}\alpha^{\prime4},\label{eq:G10-definition}
\end{equation}
we have 
\begin{equation}
\widetilde{G}_{N}^{\left(10\right)}=G_{N}^{\left(10\right)}\frac{\alpha^{\prime}}{R_{w}^{2}}.\label{eq:G10-transform}
\end{equation}
The coordinate reference volume transforms in the same way, 
\begin{equation}
\widetilde{\mathcal{V}}_{0}=L_{\widetilde{w}}V_{T^{4}}\Omega_{3}=\mathcal{V}_{0}\frac{\alpha^{\prime}}{R_{w}^{2}},\label{eq:V0-transform}
\end{equation}
and therefore Eq. (\ref{eq:global-invariant-intro}) follows.

For the backgrounds considered here, $g_{wi}=B_{wi}=0$ for every
spatial direction $i\neq w$ on the constant-time horizon section.
The only mixed components have one time index. Consequently, the Buscher
rules replace the spatial factor $\sqrt{g_{ww}}$ in the area density
by $1/\sqrt{g_{ww}}$ and leave the other spatial directions unchanged.
Hence 
\begin{equation}
\widetilde{Y}\left(r\right)=\frac{Y\left(r\right)}{g_{ww}\left(r\right)}.\label{eq:Y-Buscher-general}
\end{equation}
Together with $e^{-2\widetilde{\phi}}=e^{-2\phi}g_{ww}$, this gives
the local identity 
\begin{equation}
e^{-2\widetilde{\phi}\left(r\right)}\widetilde{Y}\left(r\right)=e^{-2\phi\left(r\right)}Y\left(r\right).\label{eq:weighted-area-Buscher-general}
\end{equation}
This identity will be verified explicitly in both duality chains.

\section{The T-dual of F1--NS5--P and the radial F1 flow}

\label{sec:NS-dual}

In this section, we apply the Buscher rules to the F1--NS5--P background
with a radial F1 source. We then use the corresponding string charge
density/bit threads correspondence to reproduce the correct black
hole entropy.

\subsection{Buscher transformation for F1-NS5-P and radial F1 source}

\label{subsec:NS-background}

We first apply Eqs. (\ref{eq:Buscher-1})--(\ref{eq:Buscher-dilaton})
to the Type IIB F1--NS5--P background. The components that enter
the transformation are 
\begin{equation}
g_{tt}=\frac{H_{P}-2}{H_{1}},\qquad g_{tw}=\frac{H_{P}-1}{H_{1}},\qquad g_{ww}=\frac{H_{P}}{H_{1}},\qquad B_{tw}=\frac{H_{1}-1}{H_{1}}.\label{eq:NS-components-before-T}
\end{equation}
Direct substitution gives 

\begin{align}
\widetilde{g}_{tt} & =g_{tt}-\frac{g_{tw}^{2}-B_{wt}^{2}}{g_{ww}}=\frac{H_{1}-2}{H_{P}},\qquad e^{2\widetilde{\phi}}=\frac{H_{5}}{H_{P}},\label{eq:NS-dual-gtt}\\
\widetilde{g}_{tw} & =\frac{H_{1}-1}{H_{P}},\qquad\widetilde{g}_{ww}=\frac{H_{1}}{H_{P}},\qquad\widetilde{B}_{tw}=\frac{H_{P}-1}{H_{P}}=1-H_{P}^{-1}.\label{eq:NS-dual-mixed-components}
\end{align}
The signs in Eq. (\ref{eq:NS-dual-mixed-components}) follow directly
from Eqs. (\ref{eq:Buscher-1}) and (\ref{eq:Buscher-2}) with the
orientation fixed above. The complete Type IIA background is therefore
\cite{Bena:2022sge}
\begin{align}
d\widetilde{S}_{\mathrm{P-NS5-F1}}^{2} & =H_{P}^{-1}\left[-dt^{2}+d\widetilde{w}^{2}+\left(H_{1}-1\right)\left(dt+d\widetilde{w}\right)^{2}\right]+H_{5}\left(dr^{2}+r^{2}d\Omega_{3}^{2}\right)+dy_{i}dy^{i},\label{eq:IIA-F1-metric}\\
e^{2\widetilde{\phi}_{\mathrm{F1}}} & =\frac{H_{5}}{H_{P}},\qquad\widetilde{B}_{2}=(1-H_{P}^{-1})dt\wedge d\widetilde{w}+2Q_{5}\omega_{2}.\label{eq:IIA-F1-fields}
\end{align}
Thus the harmonic function $H_{P}$ that represented momentum in the
original frame now represents fundamental-string winding, while $H_{1}$
now appears in the pp-wave term. The NS5 harmonic function is unchanged.
At the level of integer charges, 
\begin{equation}
\left(n_{\mathrm{F1}},n_{\mathrm{NS5}},n_{\mathrm{P}}\right)\xrightarrow{\ T_{w}\ }\left(\widetilde{n}_{\mathrm{F1}},\widetilde{n}_{\mathrm{NS5}},\widetilde{n}_{\mathrm{P}}\right)=\left(n_{\mathrm{P}},n_{\mathrm{NS5}},n_{\mathrm{F1}}\right).\label{eq:winding-momentum-exchange}
\end{equation}

Now, let us see how the fundamental string source is transformed under
this Buscher transformation. Note that the radial source used to construct
the flow is an auxiliary probe and is not one of the F1 strings that
produces the background harmonic function. Because the source ensemble
is uniformly smeared along $w$, it is compatible with the isometry
required by the Buscher rules. A transverse T-duality leaves its F1
source unchanged. Equivalently, a perturbation $B_{tr}$ is unchanged
by Eq. (\ref{eq:Buscher-2}) because $g_{wr}=B_{wr}=0$. The current
obtained by varying this component therefore maps to the corresponding
Type IIA F1 current.

Under T-duality, the Type IIA NS--NS bulk action and the F1 source
retain the same functional form:

\begin{equation}
\widetilde{S}=\frac{1}{2\widetilde{\kappa}_{10}^{2}}\int d^{10}x\sqrt{-\widetilde{g}}e^{-2\widetilde{\phi}}\left(\widetilde{R}+4\left(\widetilde{\nabla}\widetilde{\phi}\right)^{2}-\frac{1}{12}\widetilde{H}^{2}\right)+\widetilde{S}_{\mathrm{F1}},
\end{equation}
Here the radial dilaton is normalized according to $\widetilde{\phi}_{\infty}=0$.
Accordingly, $2\widetilde{\kappa}_{10}^{2}=\left(2\pi\right)^{7}\widetilde{g}_{s}^{2}\alpha^{\prime4}=16\pi\widetilde{G}_{N}^{\left(10\right)}.$
A fundamental string (F1) couples electrically to the Kalb--Ramond
two-form $B$. A macroscopic F1-string source is described by the
Polyakov action

\begin{equation}
\widetilde{S}_{\mathrm{F1}}=-\frac{1}{4\pi\alpha^{\prime}}\int d^{2}\sigma\left(\sqrt{-\gamma}\gamma^{mn}\partial_{m}\widetilde{X}^{\mathrm{a}}\partial_{n}\widetilde{X}^{\mathrm{b}}\widetilde{g}_{\mathrm{ab}}+\epsilon^{mn}\partial_{m}\widetilde{X}^{\mathrm{a}}\partial_{n}\widetilde{X}^{\mathrm{b}}\widetilde{B}_{\mathrm{ab}}\right),\label{eq:F1 source}
\end{equation}

\noindent where $\widetilde{X}^{\mathrm{a}}=\left(X^{\mathrm{\mu}},\text{\ensuremath{\widetilde{X}^{w}}}\right)$.

\subsection{From the radial F1 source to the bit-thread flow and entropy}

We now construct the bit-thread flow from the radial F1 source obtained
after T-duality. We first choose worldsheet coordinates and an embedding
adapted to the constant-time foliation. The metric in Eq. (\ref{eq:IIA-F1-metric})
has the ADM shift
\begin{equation}
\widetilde{\beta}^{\widetilde{w}}=\frac{\widetilde{g}_{t\widetilde{w}}}{\widetilde{g}_{\widetilde{w}\widetilde{w}}}=\frac{H_{1}-1}{H_{1}}.\label{eq:NS-dual-shift}
\end{equation}

\noindent To make the probe worldsheet tangent to the ADM normal on
the $t=0$ slice, we use 
\begin{equation}
X^{t}=\tau,\qquad X^{r}=\sigma,\qquad X^{\widetilde{w}}=\widetilde{w}_{0}-\frac{H_{1}\left(\sigma\right)-1}{H_{1}\left(\sigma\right)}\tau,\label{eq:NS-dual-slice-embedding}
\end{equation}
with the $S^{3}$ and $T^{4}$ coordinates held fixed. At $\tau=0$,
the induced metric satisfies 
\begin{equation}
\gamma_{\tau\tau}^{{\rm ind}}=-\frac{1}{H_{1}H_{P}},\qquad\gamma_{\tau\sigma}^{{\rm ind}}=0,\qquad\gamma_{\sigma\sigma}^{{\rm ind}}=H_{5},\qquad\det\gamma^{{\rm ind}}=-\frac{H_{5}}{H_{1}H_{P}}<0.\label{eq:NS-dual-induced-metric}
\end{equation}
The worldsheet is therefore timelike at every exterior point. Its
determinant is symmetric under $H_{1}\leftrightarrow H_{P}$, as expected
from the winding--momentum exchange. This embedding specifies the
slice data needed for the entropy calculation. Only the current at
$\tau=0$ is used, and no stationary extension of the same functional
form is assumed.

Similar to our previous calculation, after performing the $\tau$
and $\sigma$ integrations, the nonvanishing component of the antisymmetric
current (\ref{eq:F1 source}) is

\begin{equation}
\widetilde{j}_{\mathrm{F1}}^{0r}=\frac{\sqrt{\bar{\gamma}}}{\sqrt{-\widetilde{g}}}\delta\left(\widetilde{w}-\widetilde{w}_{0}\right)\delta_{T^{4}}^{\left(4\right)}\left(y-y_{0}\right)\delta_{S^{3}}\left(\Omega,\Omega_{0}\right),
\end{equation}
We then distribute $\mathcal{N}$ such radial strings uniformly over
\begin{equation}
\widetilde{\Sigma}_{8}=S_{\widetilde{w}}^{1}\times S^{3}\times T^{4},\qquad\widetilde{\mathcal{V}}_{0}=L_{\widetilde{w}}V_{T^{4}}\Omega_{3}.\label{eq:NS-dual-transverse-space}
\end{equation}
Each radial string pierces one of $\mathcal{N}$ equal cells of the
horizon section,

\begin{equation}
\Delta\widetilde{\mathcal{V}}_{0}=\frac{\widetilde{\mathcal{V}}_{0}}{\mathcal{N}}.
\end{equation}
Let the transverse coordinates be denoted collectively by $\widetilde{\xi}=\left(\widetilde{w},y^{i},\Omega^{\alpha}\right)$,
introduce the reference measure

\begin{equation}
d\mu_{0}\left(\widetilde{\xi}\right)=\sqrt{\bar{\gamma}\left(\Omega\right)}d\widetilde{w}d^{4}yd^{3}\Omega,
\end{equation}

\noindent and define 

\noindent 
\begin{equation}
\Delta_{8}\left(\widetilde{\xi},\widetilde{\xi}_{n}\right)\equiv\delta\left(\widetilde{w}-\widetilde{w}_{n}\right)\delta_{T^{4}}^{\left(4\right)}\left(y-y_{n}\right)\delta_{S^{3}}\left(\Omega,\Omega_{n}\right),
\end{equation}
which is normalized according to

\begin{equation}
\int_{\widetilde{\Sigma}_{8}}d\mu_{0}\left(\widetilde{\xi}\right)\Delta_{8}\left(\widetilde{\xi},\widetilde{\xi}_{n}\right)=1.
\end{equation}

\noindent For the $n$-th radial string, the only non-vanishing component
of the Kalb--Ramond current is

\begin{equation}
\widetilde{j}_{\mathrm{F1}}^{0r}=\frac{\sqrt{\bar{\gamma}}}{\sqrt{-\widetilde{g}}}\Delta_{8}\left(\widetilde{\xi},\widetilde{\xi}_{n}\right).
\end{equation}

\noindent Consequently, the current carried by $\mathcal{N}$ strings
is

\begin{equation}
\widetilde{j}_{\mathrm{F1}}^{0r}=\frac{\sqrt{\bar{\gamma}}}{\sqrt{-\widetilde{g}}}\sum_{n=1}^{\mathcal{N}}\Delta_{8}\left(\widetilde{\xi},\widetilde{\xi}_{n}\right).
\end{equation}

\noindent To define the continuum limit, divide $\widetilde{\Sigma}_{8}$
into $\mathcal{N}$ cells of equal reference volume $\Delta\widetilde{\mathcal{V}}_{0}=\widetilde{\mathcal{V}}_{0}/\mathcal{N}$
and place one string in each cell. The corresponding Riemann sum obeys

\begin{equation}
\Delta\widetilde{\mathcal{V}}_{0}\sum_{n=1}^{\mathcal{N}}\Delta_{8}\left(\widetilde{\xi},\widetilde{\xi}_{n}\right)\longrightarrow1.
\end{equation}
Thus, in the distributional continuum limit, the string charge density
on the $t=0$ slice is
\begin{equation}
\widetilde{j}_{\mathrm{F1}}^{0r}=\frac{\sqrt{\bar{\gamma}}}{\sqrt{-\widetilde{g}}}\frac{\mathcal{N}}{\widetilde{\mathcal{V}}_{0}}.\label{eq:NS-dual-smeared-current}
\end{equation}
Its projected string charge density is 
\begin{equation}
\widetilde{q}_{\mathrm{F1}}^{r}=\frac{\sqrt{-\widetilde{g}}}{\sqrt{\widetilde{h}}}\widetilde{j}_{\mathrm{F1}}^{0r},\qquad\partial_{I}\left(\sqrt{\widetilde{h}}\widetilde{q}_{\mathrm{F1}}^{I}\right)=0.\label{eq:NS-dual-projected-current}
\end{equation}

\noindent The induced spatial measure obtained from Eq. (\ref{eq:IIA-F1-metric})
is 

\noindent 
\begin{equation}
\sqrt{\widetilde{h}}=\sqrt{\widetilde{g}_{rr}}\widetilde{Y}_{\mathrm{F1}}\left(r\right)\sqrt{\bar{\gamma}},\qquad\widetilde{Y}_{\mathrm{F1}}\left(r\right)=\sqrt{\widetilde{g}_{\widetilde{w}\widetilde{w}}}\left(\sqrt{\widetilde{g}_{S^{3}}}\right)^{3}\sqrt{\widetilde{g}_{T^{4}}}=r^{3}H_{P}^{-1/2}H_{5}^{3/2}H_{1}^{1/2}.\label{eq:NS-dual-Y}
\end{equation}

\noindent The original Type IIB density was 

\noindent 
\begin{equation}
Y_{\mathrm{F1}}\left(r\right)=r^{3}H_{1}^{-1/2}H_{5}^{3/2}H_{P}^{1/2}.\label{eq:NS-original-Y}
\end{equation}
Since $g_{ww}=H_{P}/H_{1}$, the two are related by 
\begin{equation}
\widetilde{Y}_{\mathrm{F1}}=\frac{Y_{\mathrm{F1}}}{g_{ww}},\label{eq:NS-Y-transform-check}
\end{equation}
in agreement with Eq. (\ref{eq:Y-Buscher-general}). Define the T-dual
bit-thread flow by 
\begin{equation}
\widetilde{v}_{\mathrm{F1}}^{r}=\widetilde{C}_{{\rm geom}}^{\mathrm{F1}}\widetilde{q}_{\mathrm{F1}}^{r}=\frac{\widetilde{C}_{{\rm geom}}^{\mathrm{F1}}}{\sqrt{\widetilde{g}_{rr}}\widetilde{Y}_{\mathrm{F1}}\left(r\right)}\frac{\mathcal{N}}{\widetilde{\mathcal{V}}_{0}}.\label{eq:NS-dual-flow}
\end{equation}
Its norm is 
\begin{equation}
\left|\widetilde{v}_{\mathrm{F1}}\left(r\right)\right|=\frac{\widetilde{C}_{{\rm geom}}^{\mathrm{F1}}}{\widetilde{Y}_{\mathrm{F1}}\left(r\right)}\frac{\mathcal{N}}{\widetilde{\mathcal{V}}_{0}}.\label{eq:NS-dual-flow-norm}
\end{equation}
Using $e^{-2\widetilde{\phi}_{\mathrm{F1}}}=H_{P}/H_{5}$, we find
\begin{align}
e^{-2\widetilde{\phi}_{\mathrm{F1}}}\widetilde{Y}_{\mathrm{F1}} & =r^{3}\sqrt{H_{1}H_{5}H_{P}}\nonumber \\
 & =\sqrt{\left(r^{2}+Q_{1}\right)\left(r^{2}+Q_{5}\right)\left(r^{2}+Q_{P}\right)}.\label{eq:NS-dual-weighted-area}
\end{align}
For $Q_{1},Q_{5},Q_{P}>0$, this function is nondecreasing for $r\geq0$.
Saturating the string-frame bound at the extremal horizon therefore
fixes 
\begin{equation}
\widetilde{C}_{{\rm geom}}^{\mathrm{F1}}=\frac{\widetilde{\mathcal{V}}_{0}}{4\widetilde{G}_{N}^{\left(10\right)}\mathcal{N}}\sqrt{Q_{1}Q_{5}Q_{P}}.\label{eq:NS-dual-Cgeom}
\end{equation}
The complete pointwise check is 
\begin{equation}
\frac{\left|\widetilde{v}_{\mathrm{F1}}\left(r\right)\right|}{e^{-2\widetilde{\phi}_{\mathrm{F1}}}/\left(4\widetilde{G}_{N}^{\left(10\right)}\right)}=\left[\frac{Q_{1}Q_{5}Q_{P}}{\left(r^{2}+Q_{1}\right)\left(r^{2}+Q_{5}\right)\left(r^{2}+Q_{P}\right)}\right]^{1/2}\leq1,\label{eq:NS-dual-pointwise-bound}
\end{equation}
with equality only as $r\to0^{+}$. Finally, the flux through a constant-$r$
section is 
\begin{align}
\widetilde{\Phi}_{\mathrm{F1}} & =\int_{\widetilde{\Sigma}_{8}}\widetilde{v}_{\mathrm{F1}}^{r}\widetilde{n}_{r}d\widetilde{A}=\widetilde{C}_{{\rm geom}}^{\mathrm{F1}}\mathcal{N}\nonumber \\
 & =\frac{L_{\widetilde{w}}V_{T^{4}}\Omega_{3}}{4\widetilde{G}_{N}^{\left(10\right)}}\sqrt{Q_{1}Q_{5}Q_{P}}=\widetilde{S}_{\mathrm{BH}}^{\mathrm{P-NS5-F1}}.\label{eq:NS-dual-entropy-flux}
\end{align}
Thus, after fixing the geometric normalization by horizon saturation,
the maximal flux of the charge density derived from the T-dual F1
worldsheet equals the Bekenstein-Hawking entropy of the compactified
three-charge black hole evaluated on the full ten-dimensional horizon
section. The nontrivial point of this calculation is the T-duality
covariance of the normalized current and its capacity ratio, rather
than an independent derivation of the area law.

Moreover, let $V_{T^{4}}=\left(2\pi\right)^{4}V_{4}$. In the original
NS--NS frame, the charge parameters are \cite{Bena:2022sge,Maldacena:1996ix}
\begin{equation}
Q_{1}=\frac{g_{s}^{2}\alpha^{\prime3}}{V_{4}}n_{\mathrm{F1}},\qquad Q_{5}=\alpha^{\prime}n_{\mathrm{NS5}},\qquad Q_{P}=\frac{g_{s}^{2}\alpha^{\prime4}}{R_{w}^{2}V_{4}}n_{\mathrm{P}}.\label{eq:NS-original-charge-dictionary}
\end{equation}
In the Type IIA frame, 
\begin{equation}
\widetilde{Q}_{\mathrm{F1}}=\frac{\widetilde{g}_{s}^{2}\alpha^{\prime3}}{V_{4}}\widetilde{n}_{\mathrm{F1}},\qquad\widetilde{Q}_{\mathrm{NS5}}=\alpha^{\prime}\widetilde{n}_{\mathrm{NS5}},\qquad\widetilde{Q}_{\mathrm{P}}=\frac{\widetilde{g}_{s}^{2}\alpha^{\prime4}}{R_{\widetilde{w}}^{2}V_{4}}\widetilde{n}_{\mathrm{P}}.\label{eq:NS-dual-charge-dictionary}
\end{equation}
Eqs. (\ref{eq:T-dual-moduli}) and (\ref{eq:winding-momentum-exchange})
give 
\begin{equation}
\widetilde{Q}_{\mathrm{F1}}=Q_{P},\qquad\widetilde{Q}_{\mathrm{NS5}}=Q_{5},\qquad\widetilde{Q}_{\mathrm{P}}=Q_{1}.\label{eq:NS-charge-parameter-exchange}
\end{equation}
Substituting the dual charge dictionary into Eq. (\ref{eq:NS-dual-entropy-flux})
gives 
\begin{equation}
\widetilde{S}_{\mathrm{BH}}^{\mathrm{P-NS5-F1}}=2\pi\sqrt{\widetilde{n}_{\mathrm{F1}}\widetilde{n}_{\mathrm{NS5}}\widetilde{n}_{\mathrm{P}}}=2\pi\sqrt{n_{\mathrm{F1}}n_{\mathrm{NS5}}n_{\mathrm{P}}}.\label{eq:NS-integer-entropy}
\end{equation}
The entropy is therefore invariant even though the microscopic interpretation
of two charges has been exchanged.

\section{The T-dual of D1--D5--P and the wrapped-D2 flow}

\label{sec:RR-dual}

In this section, we apply the Buscher rules to the D1--D5--P background,
under which the radial D1 source is mapped to a D2-brane wrapped around
the dual circle. Contracting the spatially projected D2 current with
the normalized one-form along the wrapped direction yields an effective
one-dimensional radial flow. We then use the corresponding string
charge density/bit threads correspondence to reproduce the correct
black hole entropy.

\subsection{Buscher transformation for D1--D5--P background}

\label{subsec:D0D4F1-background}

For the Type IIB D1--D5--P metric, the components in the $\left(t,w\right)$
sector are 
\begin{equation}
g_{tt}=\frac{H_{P}-2}{\sqrt{H_{1}H_{5}}},\qquad g_{tw}=\frac{H_{P}-1}{\sqrt{H_{1}H_{5}}},\qquad g_{ww}=\frac{H_{P}}{\sqrt{H_{1}H_{5}}},\qquad B_{tw}=0.\label{eq:RR-components-before-T}
\end{equation}
The NS--NS Buscher rules give 
\begin{equation}
\widetilde{g}_{tt}=-\frac{1}{H_{P}\sqrt{H_{1}H_{5}}},\qquad\widetilde{g}_{tw}=0,\qquad\widetilde{g}_{ww}=\frac{\sqrt{H_{1}H_{5}}}{H_{P}},\qquad\widetilde{B}_{tw}=1-H_{P}^{-1}.\label{eq:RR-dual-components}
\end{equation}
The transformed dilaton is 
\begin{equation}
e^{2\widetilde{\phi}_{\mathrm{D2}}}=\frac{e^{2\phi_{\mathrm{D1}}}}{g_{ww}}=H_{1}^{3/2}H_{5}^{-1/2}H_{P}^{-1}.\label{eq:D0D4F1-dilaton}
\end{equation}

The R--R potentials transform according to the standard Type II T-duality
rules \cite{Hassan:1999bv,Myers:1999ps}. The electric component of
$C_{2}$ with one $w$ index becomes a Type IIA one-form, while the
magnetic component without a $w$ index becomes a three-form with
one $\widetilde{w}$ index. Specifically, let $\mu,\nu,\ldots$ denote
the nine coordinates other than the duality direction $w$. Up to
an overall sign fixed by the R--R convention and the orientation
of the dual circle, the massless Type II rules can be written as

\begin{align}
\widetilde{C}_{\widetilde{w}\mu_{2}\cdots\mu_{n}}^{\left(n\right)} & =a_{\left(A-B\right)}\left[C_{\mu_{2}\cdots\mu_{n}}^{\left(n-1\right)}-\left(n-1\right)g_{ww}^{-1}g_{w\left[\mu_{2}\right.}C_{\left.w\mu_{3}\cdots\mu_{n}\right]}^{\left(n-1\right)}\right],\label{eq:RR-T-duality-with-w}\\
\widetilde{C}_{\mu_{1}\cdots\mu_{n}}^{\left(n\right)} & =a_{\left(A-B\right)}C_{w\mu_{1}\cdots\mu_{n}}^{\left(n+1\right)}-nB_{w\left[\mu_{1}\right.}\widetilde{C}_{\left.\widetilde{w}\mu_{2}\cdots\mu_{n}\right]}^{\left(n\right)}.\label{eq:RR-T-duality-without-w}
\end{align}
Here $a_{\left(A-B\right)}=\pm1$ is convention dependent. We choose
$a_{\left(A-B\right)}$, together with the orientation of $\widetilde{w}$,
such that a component with one $w$ index loses that index with a
positive sign, whereas a component without a $w$ index acquires a
$d\widetilde{w}$ factor with a positive sign. Schematically, due
to $B_{w\mu}=0$ and $g_{wt}\neq0$ in our solution, we have

\begin{equation}
C_{\mu_{1}\cdots\mu_{p-1}w}^{\left(p\right)}\longrightarrow\widetilde{C}_{\mu_{1}\cdots\mu_{p-1}}^{\left(p-1\right)},\qquad C_{\mu_{1}\cdots\mu_{p}}^{\left(p\right)}\longrightarrow\widetilde{C}_{\mu_{1}\cdots\mu_{p}\widetilde{w}}^{\left(p+1\right)}.\label{eq:RR-T-duality-schematic}
\end{equation}
The first operation contracts the original form with the Killing vector
$\partial_{w}$, while the second adds the dual-circle one-form. This
converts the even-degree R--R potentials of Type IIB into the odd-degree
potentials of Type IIA.

In the Type IIB D1--D5--P background, we use the gauge 

\begin{equation}
C^{\left(2\right)}=\left(1-H_{1}^{-1}\right)dt\wedge dw+2Q_{5}\omega_{2},\qquad B_{2}=0.\label{eq:IIB-RR-potential-split}
\end{equation}

\noindent The first term in Eq. (\ref{eq:IIB-RR-potential-split})
is the electric potential sourced by the D1 charge. Since it contains
one $w$ index, it becomes a Type IIA one-form: 
\begin{equation}
C_{tw}^{\left(2\right)}=1-H_{1}^{-1}\quad\longrightarrow\quad\widetilde{C}_{t}^{\left(1\right)}=1-H_{1}^{-1}.\label{eq:Ctw-to-Ct}
\end{equation}

\noindent This is consistent with the background charge map 
\begin{equation}
\mathrm{D1}\left(w\right)\xrightarrow{\ T_{w}\ }\mathrm{D0}.\label{eq:D1-to-D0}
\end{equation}
Indeed, the original D1-brane couples electrically to $C_{tw}^{\left(2\right)}$,
whereas the T-dual D0-brane couples electrically to $\widetilde{C}_{t}^{(1)}$.

The second term in Eq. (\ref{eq:IIB-RR-potential-split}) is the magnetic
potential associated with the D5 charge. It has no $w$ index and
therefore becomes a three-form with one $\widetilde{w}$ index. For
two indices $\alpha,\beta$ on $S^{3}$, Eq. (\ref{eq:RR-T-duality-with-w})
gives 

\begin{equation}
C_{\alpha\beta}^{\left(2\right)}=2Q_{5}\left(\omega_{2}\right)_{\alpha\beta}\quad\longrightarrow\quad\widetilde{C}_{\alpha\beta\widetilde{w}}^{\left(3\right)}=2Q_{5}\left(\omega_{2}\right)_{\alpha\beta}\wedge d\widetilde{w},
\end{equation}

\noindent or, equivalently, $\widetilde{C}^{(3)}=2Q_{5}\omega_{2}\wedge d\widetilde{w}$.
This agrees with 
\begin{equation}
\mathrm{D5}\left(wT^{4}\right)\xrightarrow{\ T_{w}\ }\mathrm{D4}\left(T^{4}\right).\label{eq:D5-to-D4}
\end{equation}
The D5-brane is magnetically charged under $C^{\left(2\right)}$,
whereas the T-dual D4-brane is magnetically charged under $\widetilde{C}^{\left(3\right)}$.
Accordingly, the original magnetic flux through $S^{3}$ becomes a
four-form flux through $S^{3}\times S_{\widetilde{w}}^{1}$. Because
the IIB truncation also has $C^{\left(4\right)}=0$, no additional
components of $\widetilde{C}^{\left(3\right)}$ are generated.

We thus obtain the following consistent gauge choice for the R-R potentials
in this truncation:

\begin{equation}
\widetilde{C}^{\left(1\right)}=\left(1-H_{1}^{-1}\right)dt,\qquad\widetilde{C}^{\left(3\right)}=2Q_{5}\omega_{2}\wedge d\widetilde{w}.\label{eq:D0D4F1-RR-potentials}
\end{equation}
The full Type IIA solution is therefore \cite{Bena:2008dw}
\begin{align}
d\widetilde{S}_{\mathrm{D0-D4-F1}}^{2} & =-H_{1}^{-1/2}H_{5}^{-1/2}H_{P}^{-1}dt^{2}+H_{1}^{1/2}H_{5}^{1/2}H_{P}^{-1}d\widetilde{w}^{2}\nonumber \\
 & \quad+\left(H_{1}H_{5}\right)^{1/2}\left(dr^{2}+r^{2}d\Omega_{3}^{2}\right)+\left(\frac{H_{1}}{H_{5}}\right)^{1/2}dy_{i}dy^{i},\label{eq:D0D4F1-metric}\\
\widetilde{B}^{\left(2\right)} & =\left(1-H_{P}^{-1}\right)dt\wedge d\widetilde{w},\qquad e^{2\widetilde{\phi}_{\mathrm{D2}}}=H_{1}^{3/2}H_{5}^{-1/2}H_{P}^{-1},\label{eq:D0D4F1-NS-fields}\\
\widetilde{C}^{\left(1\right)} & =\left(1-H_{1}^{-1}\right)dt,\qquad\widetilde{C}^{\left(3\right)}=2Q_{5}\omega_{2}\wedge d\widetilde{w}.\label{eq:D0D4F1-fields-repeat}
\end{align}
Here $H_{1}$ carries D0 charge, $H_{5}$ carries D4 charge, and $H_{P}$
carries F1 winding charge. The D4-branes wrap $T^{4}$ but not $\widetilde{w}$.
The D0 and D4 sources are uniformly smeared along $S_{\widetilde{w}}^{1}$
in the supergravity zero-mode description. The brane map is 
\begin{equation}
\mathrm{D1}\left(w\right)\to\mathrm{D0},\qquad\mathrm{D5}\left(wT^{4}\right)\to\mathrm{D4}\left(T^{4}\right),\qquad\mathrm{P}\left(w\right)\to\mathrm{F1}\left(\widetilde{w}\right).\label{eq:D-brane-background-map}
\end{equation}
The corresponding Type IIA string-frame bulk action can be written
as 
\begin{align}
S_{\mathrm{IIA}} & =\frac{1}{2\widetilde{\kappa}_{10}^{2}}\int d^{10}x\sqrt{-\widetilde{g}}\left\{ e^{-2\widetilde{\phi}}\left[\widetilde{R}+4\left(\partial\widetilde{\phi}\right)^{2}-\frac{1}{12}\left(\widetilde{H}^{\left(3\right)}\right)^{2}\right]-\frac{1}{4}\left(\widetilde{F}^{\left(2\right)}\right)^{2}-\frac{1}{48}\left(\widetilde{F}^{\left(4\right)}\right)^{2}\right\} +S_{{\rm CS}},\label{eq:IIA-action}\\
\widetilde{H}^{\left(3\right)} & =d\widetilde{B}^{\left(2\right)},\qquad\widetilde{F}^{\left(2\right)}=d\widetilde{C}^{\left(1\right)},\qquad\widetilde{F}^{\left(4\right)}=d\widetilde{C}^{\left(3\right)}-\widetilde{C}^{\left(1\right)}\wedge\widetilde{H}^{\left(3\right)}.\label{eq:IIA-field-strengths}
\end{align}
In Eq. (\ref{eq:IIA-action}), the constant asymptotic dilaton factor
is absorbed into the ten-dimensional gravitational coupling, so that
$2\widetilde{\kappa}_{10}^{2}=\left(2\pi\right)^{7}\widetilde{g}_{s}^{2}\alpha^{\prime4}=16\pi\widetilde{G}_{N}^{\left(10\right)}$.
 The field $\widetilde{\phi}$ denotes the normalized radial dilaton
satisfying $\widetilde{\phi}_{\infty}=0$. For the background above,
$\widetilde{C}^{\left(1\right)}\wedge\widetilde{H}_{3}=0$ because
both electric forms contain $dt$. The Chern--Simons term does not
affect the radial current calculation below.

\subsection{T-duality of the radial D1-brane source}

\label{subsec:D1-to-D2-source}

We begin with the component form of the D1 source used in our previous
work, because it makes the connection with the worldsheet current
explicit: 

\begin{equation}
S_{\mathrm{D1}}=-\frac{1}{4\pi\alpha^{\prime}}\int d^{2}\sigma\left(e^{-\phi}\sqrt{-\gamma}\gamma^{mn}\partial_{m}X^{\mathrm{a}}\partial_{n}X^{\mathrm{b}}g_{\mathrm{ab}}-\epsilon^{mn}\partial_{m}X^{\mathrm{a}}\partial_{n}X^{\mathrm{b}}C_{\mathrm{ab}}^{\left(2\right)}\right),\label{eq:previous-D1-component-action}
\end{equation}

\noindent where the second Wess--Zumino part is denoted by $S_{\mathrm{D1}}^{\mathrm{WZ}}$.
We choose the worldsheet orientation such that the relative minus
sign inside the parentheses gives the positive R-R coupling. Reversing
the worldsheet orientation reverses the sign of the D1 charge.

Now we use the pullback notation to simplify the following calculations.
Varying the action with respect to the inverse auxiliary worldsheet
metric $\gamma^{mn}$, while keeping the embedding fields and the
spacetime background fixed, gives
\begin{equation}
P\left[g\right]_{mn}-\frac{1}{2}\gamma_{mn}\gamma^{pq}P\left[g\right]_{pq}=0,\qquad P\left[g\right]_{mn}\equiv\partial_{m}X^{\mathrm{a}}\partial_{n}X^{\mathrm{b}}g_{\mathrm{ab}}.\label{eq:D1-auxiliary-metric-equation}
\end{equation}
In two dimensions this equation fixes $\gamma_{mn}$ to be conformal
to the induced metric. Using worldsheet Weyl invariance, one may set
$\gamma_{mn}=P\left[g\right]_{mn}$. Equation (\ref{eq:previous-D1-component-action})
then becomes 
\begin{equation}
S_{\mathrm{D1}}=-\frac{1}{2\pi\alpha^{\prime}}\int d^{2}\sigma e^{-\phi}\sqrt{-\det P\left[g\right]}+\frac{1}{4\pi\alpha^{\prime}}\int d^{2}\sigma\epsilon^{mn}P\left[C^{\left(2\right)}\right]_{mn},\label{eq:D1-DBI-WZ-truncated}
\end{equation}
with $P\left[C^{\left(2\right)}\right]_{mn}\equiv\partial_{m}X^{\mathrm{a}}\partial_{n}X^{\mathrm{b}}C_{\mathrm{ab}}^{\left(2\right)}$.
Here $\det P\left[g\right]$ denotes the determinant of the two-dimensional
induced metric, not that of the ten-dimensional spacetime metric.

\begin{figure}[h]
\begin{centering}
\includegraphics[scale=0.35]{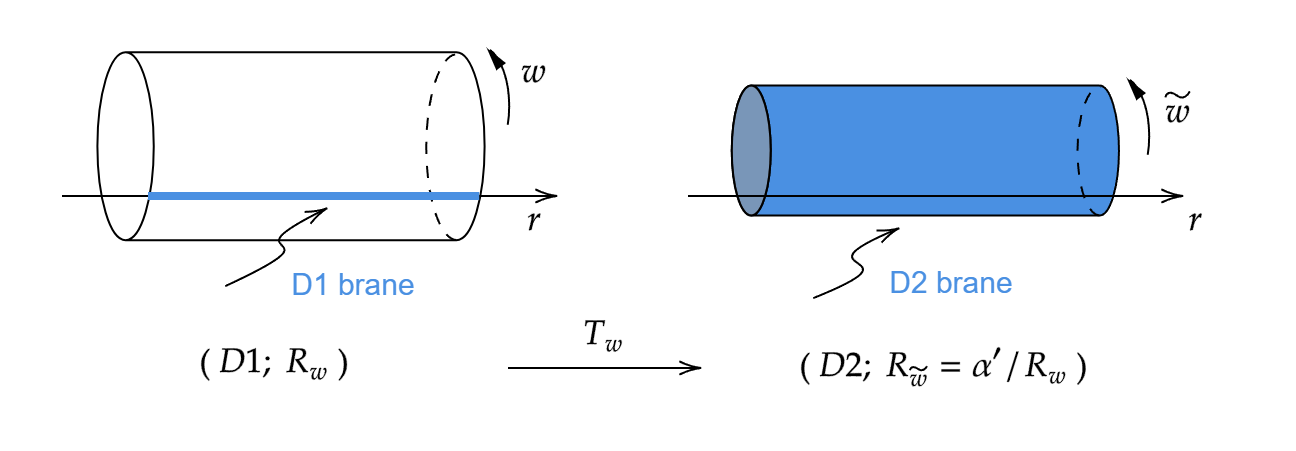}
\par\end{centering}
\caption{\label{fig:D1T}This figure illustrates a D1-brane extended along
the radial direction $r$ and localized on the compact $w$ circle.
Under T-duality along $w$, it maps to a D2-brane extended along $r$
and wrapped on the dual circle $\widetilde{w}$.}
\end{figure}

The radial D1 has worldvolume directions $\left(t,r\right)$, whereas
$w$ is a transverse direction. T-duality along $w$ therefore gives
\begin{equation}
\mathrm{D1}\left(t,r\right)\xrightarrow{\;T_{w}\;}\mathrm{D2}\left(t,r,\widetilde{w}\right).\label{eq:radial-D1-to-D2}
\end{equation}
The T-duality map between the radial D1-brane and the wrapped D2-brane
is illustrated in Fig. (\ref{fig:D1T}).
 The dual worldvolume has the product form 
\begin{equation}
\mathcal{W}_{3}=\mathcal{W}_{2}\times S_{\widetilde{w}}^{1}.\label{eq:D2-worldvolume-product-source}
\end{equation}
Let $\mu,\nu$ label the spacetime directions other than $w$. Split
the ten-dimensional index as $\mathrm{a}=\left(\mu,w\right)$, $\mu\neq w$.
The pullback of the Type IIB two-form (\ref{eq:D1-DBI-WZ-truncated})
can be decomposed as 

\begin{equation}
P\left[C^{\left(2\right)}\right]_{mn}=C_{\mu\nu}^{\left(2\right)}\partial_{m}X^{\mu}\partial_{n}X^{\nu}+2C_{\mu w}^{\left(2\right)}\partial_{\left[m\right.}X^{\mu}\partial_{\left.n\right]}X^{w}.\label{eq:D1-C2-pullback-decomposition}
\end{equation}
Schematically, the R--R T-duality rules (\ref{eq:RR-T-duality-schematic})
give 

\begin{equation}
C_{\mu\nu}^{\left(2\right)}\longrightarrow\widetilde{C}_{\mu\nu\widetilde{w}}^{\left(3\right)},\qquad C_{\mu w}^{\left(2\right)}\longrightarrow\widetilde{C}_{\mu}^{\left(1\right)}.\label{eq:RR-components-D1-to-D2}
\end{equation}
The first term $C_{\mu\nu}^{\left(2\right)}\partial_{m}X^{\mu}\partial_{n}X^{\nu}$
in Eq. (\ref{eq:D1-C2-pullback-decomposition}) therefore becomes
the D2 coupling to $\widetilde{C}_{3}$:

\begin{equation}
P\left[\widetilde{C}^{\left(3\right)}\right]_{mnp}=\widetilde{C}_{\mu\nu\widetilde{w}}^{\left(3\right)}\partial_{m}X^{\mu}\partial_{n}X^{\nu}\partial_{p}X^{\widetilde{w}}.
\end{equation}
In the second term $2C_{\mu w}^{\left(2\right)}\partial_{\left[m\right.}X^{\mu}\partial_{\left.n\right]}X^{w}$,
the coupling $C_{\mu w}^{\left(2\right)}$ becomes $\widetilde{C}_{\mu}^{\left(1\right)}$.
Moreover, the transverse T-duality transforms the original scalar
$X^{w}$ to the Wilson-line component of the D2 gauge field, 
\begin{equation}
X^{w}=2\pi\alpha^{\prime}A_{\widetilde{w}}.\label{eq:scalar-Wilson-line-map}
\end{equation}
It is worth noting that in our case, a D1-brane localized at a constant
position $X^{w}=w_{0}$ on the $t=0$ hypersurface is mapped to a
flat D2 connection $A_{\widetilde{w}}=w_{0}/\left(2\pi\alpha^{\prime}\right)$,
whose Wilson line records the original D1 position. For a general
worldvolume-dependent transverse scalar and fields independent of
$\widetilde{w}$, this implies 
\begin{equation}
\partial_{m}X^{w}=2\pi\alpha^{\prime}F_{m\widetilde{w}},\qquad m=0,1.\label{eq:scalar-derivative-gauge-field}
\end{equation}
The second term in Eq. (\ref{eq:D1-C2-pullback-decomposition}) therefore
produces $\widetilde{C}_{\mathrm{a}}^{\left(1\right)}\partial_{m}X^{\mathrm{a}}\left(2\pi\alpha^{\prime}F_{n\widetilde{w}}\right)$.
However, the complete dual coupling must be expressed in terms of
the gauge-invariant combination

\begin{equation}
\mathcal{F}_{m\widetilde{w}}=P\left[\widetilde{B}^{\left(2\right)}\right]_{m\widetilde{w}}+2\pi\alpha^{\prime}F_{m\widetilde{w}},
\end{equation}
where $\widetilde{B}^{\left(2\right)}$ is obtained by the NS-NS Buscher
rules and worldvolume gauge invariance. 

For the source used in our construction, the worldvolume gauge field
is fixed more explicitly by the T-dual of the normal-aligned D1 embedding.
In the original D1-D5-P frame, the ADM shift is
\begin{equation}
\beta^{w}\left(r\right)=\frac{g_{tw}}{g_{ww}}=\frac{H_{P}-1}{H_{P}},
\end{equation}
 and the D1 embedding on the selected foliation is
\begin{equation}
X^{t}=\tau,\qquad X^{r}=\sigma,\qquad X^{w}=w_{0}-\beta^{w}\left(r\right)t.
\end{equation}
 The T-dual map therefore gives 
\begin{equation}
2\pi\alpha^{\prime}F_{t\widetilde{w}}=-\frac{H_{P}-1}{H_{P}},\qquad2\pi\alpha^{\prime}F_{r\widetilde{w}}=-t\partial_{r}\left(\frac{H_{P}-1}{H_{P}}\right).
\end{equation}
 Since the T-dual background has
\begin{equation}
\widetilde{B}_{t\widetilde{w}}=\frac{H_{P}-1}{H_{P}},
\end{equation}
 the gauge-invariant field strength satisfies
\begin{equation}
\mathcal{F}_{t\widetilde{w}}=P\left[\widetilde{B}^{\left(2\right)}\right]_{t\widetilde{w}}+2\pi\alpha^{\prime}F_{t\widetilde{w}}=0,\qquad\left.\mathcal{F}_{r\widetilde{w}}\right|_{t=0}=0.
\end{equation}
Thus the D2 action is Lorentzian on the $t=0$ slice and is the direct
T-dual of the normal D1 worldvolume. On this slice, the induced coupling
$P\left[\widetilde{C}^{\left(1\right)}\right]\wedge\mathcal{F}$ also
vanishes. 

Although $\mathcal{F}$ is required in the complete D2 DBI and Wess-Zumino
actions, it does not modify the D2 three-form current, because that
current is defined by variation with respect to the independent potential
$\widetilde{C}^{\left(3\right)}$. Variation with respect to $\widetilde{B}^{\left(2\right)}$
instead defines a distinct NS-NS fundamental-string current carried
by the D2 worldvolume, which is not part of the wrapped-D2 charge
flow studied here. Combining the two R-R contributions, the dual D2
Wess-Zumino action takes the form

\begin{equation}
S_{\mathrm{D1}}^{\mathrm{WZ}}\xrightarrow{\;T_{w}\;}S_{\mathrm{D2}}^{\mathrm{WZ}}=\frac{1}{3!\left(2\pi\right)^{2}\alpha^{\prime3\text{/2}}}\int d^{3}\xi\epsilon^{mnp}\left[\partial_{m}X^{\mathrm{a}}\partial_{n}X^{\mathrm{b}}\partial_{p}X^{\mathrm{c}}\widetilde{C}_{\mathrm{abc}}+3\partial_{m}X^{\mathrm{a}}\widetilde{C}_{\mathrm{a}}\mathcal{F}_{np}\right].\label{eq:complete-D2-WZ-components}
\end{equation}
The D2 current relevant to our correspondence is defined by varying
the full D2 source action with respect to $\widetilde{C}_{3}$. Since
the second term is independent of $\widetilde{C}_{3}$, it does not
contribute to the D2 three-form current. Instead, its variation with
respect to $\widetilde{C}_{1}$ defines a separate induced D0 current.
We may therefore isolate the $\widetilde{C}_{3}$ coupling,

\begin{equation}
S_{\mathrm{D2}}^{\mathrm{WZ}}=\frac{1}{3!\left(2\pi\right)^{2}\alpha^{\prime3\text{/2}}}\int d^{3}\xi\epsilon^{mnp}\partial_{m}X^{\mathrm{a}}\partial_{n}X^{\mathrm{b}}\partial_{p}X^{\mathrm{c}}\widetilde{C}_{\mathrm{abc}}\left(X\right),
\end{equation}
which is consistent with the WZ action of a p-brane \cite{Frey:2019fqz}
and superstring source \cite{Blau:2002}. The corresponding spacetime
current is derived from this three-dimensional action before any restriction
to a spatial hypersurface is imposed. After deriving the current,
we evaluate its spatial projection on the $t=0$ slice used in the
original D1 construction. 

\subsection{From the D2-brane source to the bit-thread flow and entropy }

\label{subsec:D2-effective-current}

Now, we begin with the Wess--Zumino part of the D2 action

\begin{equation}
S_{\mathrm{D2}}^{\mathrm{WZ}}=\frac{1}{3!\left(2\pi\right)^{2}\alpha^{\prime3\text{/2}}}\int d^{3}\xi\epsilon^{mnp}\partial_{m}X^{\mathrm{a}}\partial_{n}X^{\mathrm{b}}\partial_{p}X^{\mathrm{c}}\widetilde{C}_{\mathrm{abc}}\left(X\right).
\end{equation}

\noindent Varying the $\widetilde{C}_{3}$ gives the D2 spacetime
current 
\begin{equation}
\widetilde{j}_{\mathrm{D2}}^{\mathrm{abc}}\left(x\right)=\frac{1}{\sqrt{-\widetilde{g}}}\int d^{3}\xi\epsilon^{mnp}\partial_{m}X^{\mathrm{a}}\partial_{n}X^{\mathrm{b}}\partial_{p}X^{\mathrm{c}}\delta^{\left(10\right)}\left(x-X\left(\xi\right)\right).\label{eq:D2-current-general}
\end{equation}
After T-duality, the D2-brane worldvolume extends along $\left(t,r,\widetilde{w}\right)$.
Choose worldvolume coordinates

\begin{equation}
\xi^{m}=\left(\xi^{0},\xi^{1},\xi^{2}\right)=\left(\tau,\sigma,\vartheta\right),
\end{equation}

\noindent we choose the static gauge

\begin{equation}
X^{t}\left(\xi\right)=\tau,\qquad X^{r}\left(\xi\right)=\sigma,\qquad X^{\widetilde{w}}\left(\xi\right)=\vartheta,
\end{equation}

\noindent with $X^{\alpha}=\Omega_{0}^{\alpha}$ ($\alpha=1,2,3$)
and $X^{i}=y_{0}^{i}$ ($i=1,\ldots,4$). Setting the worldvolume
orientation to $\epsilon^{\tau\sigma\vartheta}=+1$, we obtain the
nonvanishing component

\begin{eqnarray}
\widetilde{j}_{\mathrm{D2}}^{0r\widetilde{w}}\left(x\right) & = & \frac{1}{\sqrt{-\widetilde{g}}}\int d\tau d\sigma d\vartheta\delta^{\left(10\right)}\left(x-X\left(\tau,\sigma,\vartheta\right)\right)\nonumber \\
 & = & \frac{1}{\sqrt{-\widetilde{g}}}\int d\tau\delta\left(t-\tau\right)\int d\sigma\delta\left(r-\sigma\right)\int_{0}^{L_{\widetilde{w}}}d\vartheta\delta\left(\widetilde{w}-\vartheta\right)\times\nonumber \\
 &  & \delta_{T^{4}}^{\left(4\right)}\left(y-y_{0}\right)\delta_{S^{3}}\left(\Omega,\Omega_{0}\right).
\end{eqnarray}
Using the periodic delta function on $S_{\widetilde{w}}^{1}$, the
result is

\begin{equation}
\widetilde{j}_{\mathrm{D2}}^{0r\widetilde{w}}\left(x\right)=\frac{\sqrt{\bar{\gamma}}}{\sqrt{-\widetilde{g}}}\delta_{T^{4}}^{\left(4\right)}\left(y-y_{0}\right)\delta_{S^{3}}\left(\Omega,\Omega_{0}\right),
\end{equation}

\noindent where the factor $\sqrt{\bar{\gamma}}$ comes from rewriting
the coordinate delta function on $S^{3}$ as a covariant delta function.

Now, we consider $\mathcal{N}$ wrapped D2-branes located at $\left(y_{n},\Omega_{n}\right)$,
with $n=1,\ldots,\mathcal{N}$. The current carried by these branes
is

\begin{equation}
\widetilde{j}_{\mathrm{D2}}^{0r\widetilde{w}}\left(x\right)=\frac{\sqrt{\bar{\gamma}}}{\sqrt{-\widetilde{g}}}\sum_{n=1}^{\mathcal{N}}\delta_{T^{4}}^{\left(4\right)}\left(y-y_{n}\right)\delta_{S^{3}}\left(\Omega,\Omega_{n}\right),
\end{equation}
Let the transverse coordinates be denoted collectively by $\zeta=\left(y^{i},\Omega^{\alpha}\right)$,
and introduce the reference measure

\begin{equation}
d\mu_{0}\left(\zeta\right)=\sqrt{\bar{\gamma}\left(\Omega\right)}d^{4}yd^{3}\Omega.
\end{equation}

\noindent The corresponding reference volume is

\begin{equation}
\widetilde{\mathcal{V}}_{\perp}=V_{T^{4}}\Omega_{3},\qquad\Omega_{3}=2\pi^{2}.
\end{equation}

\noindent Here $V_{T^{4}}=\int d^{4}y$ denotes the coordinate reference
volume. Thus, $\widetilde{\mathcal{V}}_{\perp}$ is independent of
the dynamical string-frame metric and should not be confused with
the proper volume of $\Sigma_{7}$. Define 

\noindent 
\begin{equation}
\Delta_{7}\left(\zeta,\zeta_{n}\right)\equiv\delta_{T^{4}}^{\left(4\right)}\left(y-y_{n}\right)\delta_{S^{3}}\left(\Omega,\Omega_{n}\right),
\end{equation}

\noindent which is normalized according to

\begin{equation}
\int_{\Sigma_{7}}d\mu_{0}\left(\zeta\right)\Delta_{7}\left(\zeta,\zeta_{n}\right)=1.
\end{equation}
Summing the only nonvanishing component over all $\mathcal{N}$ branes
gives

\begin{equation}
\widetilde{j}_{\mathrm{D2}}^{0r\widetilde{w}}\left(x\right)=\frac{\sqrt{\bar{\gamma}}}{\sqrt{-\widetilde{g}}}\sum_{n=1}^{\mathcal{N}}\Delta_{7}\left(\zeta,\zeta_{n}\right),
\end{equation}

\noindent To define the continuum limit, divide $\Sigma_{7}$ into
$\mathcal{N}$ cells of equal reference volume $\Delta\widetilde{\mathcal{V}}_{\perp}=\widetilde{\mathcal{V}}_{\perp}/\mathcal{N}$
and place one brane in each cell. The corresponding Riemann sum obeys

\begin{equation}
\Delta\widetilde{\mathcal{V}}_{\perp}\sum_{n=1}^{\mathcal{N}}\Delta_{7}\left(\zeta,\zeta_{n}\right)\longrightarrow1,
\end{equation}

\noindent in the distributional sense. Therefore, the smeared current
becomes 

\begin{equation}
\widetilde{j}_{\mathrm{D2}}^{0r\widetilde{w}}\left(x\right)=\frac{\sqrt{\bar{\gamma}}}{\sqrt{-\widetilde{g}}}\frac{\mathcal{N}}{\widetilde{\mathcal{V}}_{\perp}},
\end{equation}
It obeys 
\begin{equation}
\partial_{a}\left(\sqrt{-\widetilde{g}}\widetilde{j}_{\mathrm{D2}}^{\mathrm{abc}}\right)=0.\label{eq:D2-current-conservation}
\end{equation}
Each dual D2 wraps the whole $S_{\widetilde{w}}^{1}$. Its spatial
projection is the antisymmetric bivector 
\begin{equation}
Q_{\mathrm{D2}}^{IJ}\equiv\widetilde{N}_{{\rm ADM}}\widetilde{j}_{\mathrm{D2}}^{0IJ}=\frac{\sqrt{-\widetilde{g}}}{\sqrt{\widetilde{h}}}\widetilde{j}_{\mathrm{D2}}^{0IJ},\qquad Q_{\mathrm{D2}}^{IJ}=-Q_{\mathrm{D2}}^{JI}.\label{eq:D2-projected-bivector}
\end{equation}
Because the ADM normal covector has components $n_{a}=\left(-\widetilde{N}_{{\rm ADM}},0,\ldots,0\right)$,
this projection formula is unchanged by a stationary shift. The present
D0--D4--F1 metric is in fact static, with $\widetilde{g}_{t\widetilde{w}}=0$.
In the present configuration, 
\begin{equation}
Q_{\mathrm{D2}}=Q_{\mathrm{D2}}^{r\widetilde{w}}\partial_{r}\wedge\partial_{\widetilde{w}}.\label{eq:D2-bivector-geometric}
\end{equation}

\noindent Thus the ten-dimensional spatial current is an oriented
area flow in the $\left(r,\widetilde{w}\right)$ plane, rather than
an ordinary radial vector. Therefore, this bivector cannot be identified
directly with a bit-thread vector. To extract the current seen after
reduction along the wrapped direction, we introduce the normalized
one-form 

\begin{equation}
\eta\equiv\frac{d\widetilde{w}}{L_{\widetilde{w}}},\qquad\int_{S_{\widetilde{w}}^{1}}\eta=1,\qquad d\eta=0.\label{eq:normalized-circle-one-form}
\end{equation}

\noindent Here $L_{\widetilde{w}}$ denotes the fixed coordinate period
of the dual circle, not its local proper circumference. The latter
generally depends on the radial coordinate through $\widetilde{g}_{\widetilde{w}\widetilde{w}}\left(r\right)$
and would not define the closed one-form needed below. The form $\eta$
is the normalized generator of $H^{1}\left(S_{\widetilde{w}}^{1}\right)$.
It is invariant under a constant rescaling $\widetilde{w}^{\prime}=\lambda\widetilde{w}$.
This normalization therefore removes the dependence on the arbitrary
choice of coordinate used to parametrize the circle.

The reduced spatial charge density is defined by contracting the circle
leg of the bivector with $\eta$, 
\begin{equation}
q_{\mathrm{D2}}^{I}\equiv\left(Q_{\mathrm{D2}}\cdot\eta\right)^{I}\equiv Q_{\mathrm{D2}}^{IJ}\eta_{J}.\label{eq:reduced-current-definition}
\end{equation}
This is a dimensional reduction of the wrapped-D2 charge density,
not an additional ten-dimensional string current. Since 
\begin{equation}
\eta_{J}=\frac{1}{L_{\widetilde{w}}}\delta_{J}^{\widetilde{w}},
\end{equation}
Eq. (\ref{eq:reduced-current-definition}) gives 
\begin{equation}
q_{\mathrm{D2}}^{I}=\frac{Q_{\mathrm{D2}}^{I\widetilde{w}}}{L_{\widetilde{w}}}.\label{eq:reduced-current-components}
\end{equation}
Consequently, 
\begin{equation}
q_{\mathrm{D2}}^{r}=\frac{Q_{\mathrm{D2}}^{r\widetilde{w}}}{L_{\widetilde{w}}},\qquad q_{\mathrm{D2}}^{\widetilde{w}}=\frac{Q_{\mathrm{D2}}^{\widetilde{w}\widetilde{w}}}{L_{\widetilde{w}}}=0,\qquad q^{\perp}=0,\label{eq:radial-reduced-current}
\end{equation}
where $\perp$ labels the remaining transverse directions. In other
words, the map acts geometrically as 
\begin{equation}
\partial_{r}\wedge\partial_{\widetilde{w}}\longrightarrow\frac{1}{L_{\widetilde{w}}}\partial_{r}.\label{eq:bivector-to-vector-map}
\end{equation}
An overall minus sign may appear if the opposite convention is chosen
for the interior product. Note that Eq. (\ref{eq:reduced-current-definition})
fixes the orientation convention used here.

For a stationary, circle-independent source, the spatial D2 current
obeys 
\begin{equation}
D_{I}Q_{\mathrm{D2}}^{IJ}=0,\label{eq:D2-spatial-conservation}
\end{equation}
where $D_{I}$ denotes the Levi-Civita covariant derivative associated
with the induced spatial metric $\widetilde{h}_{IJ}$ on the constant-time
hypersurface. Using the antisymmetry of $Q_{\mathrm{D2}}^{IJ}$ and
the fact that $\eta$ is closed, we find 
\begin{align}
D_{I}q_{\mathrm{D2}}^{I} & =D_{I}\left(Q_{\mathrm{D2}}^{IJ}\eta_{J}\right)\nonumber \\
 & =\left(D_{I}Q_{\mathrm{D2}}^{IJ}\right)\eta_{J}+Q_{\mathrm{D2}}^{IJ}D_{I}\eta_{J}\nonumber \\
 & =Q_{\mathrm{D2}}^{IJ}D_{\left[I\right.}\eta_{\left.J\right]}=\frac{1}{2}Q_{\mathrm{D2}}^{IJ}\left(d\eta\right)_{IJ}=0.\label{eq:reduced-current-conservation}
\end{align}
For the radial configuration this result can also be read directly
from 
\begin{equation}
\partial_{r}\left(\sqrt{\widetilde{h}}Q_{\mathrm{D2}}^{r\widetilde{w}}\right)=0,\qquad\partial_{r}\left(\sqrt{\widetilde{h}}q_{\mathrm{D2}}^{r}\right)=0.\label{eq:radial-density-conservation}
\end{equation}
Thus contraction with $\eta$ not only lowers the rank of the current
but also preserves its conservation law. Using the normalized one-form
$\eta$ in Eq. (\ref{eq:reduced-current-definition}), we have
\begin{equation}
q_{\mathrm{D2}}^{I}=Q_{\mathrm{D2}}^{IJ}\eta_{J}.\label{eq:D2-effective-vector-definition}
\end{equation}
For the radial configuration, 
\begin{equation}
q_{\mathrm{D2}}^{r}=\frac{1}{L_{\widetilde{w}}}Q_{\mathrm{D2}}^{r\widetilde{w}}=\frac{\sqrt{\bar{\gamma}}}{\sqrt{\widetilde{h}}}\frac{\mathcal{N}}{L_{\widetilde{w}}\widetilde{\mathcal{V}}_{\perp}}=\frac{\sqrt{\bar{\gamma}}}{\sqrt{\widetilde{h}}}\frac{\mathcal{N}}{\widetilde{\mathcal{V}}_{0}}.\label{eq:D2-effective-radial-current}
\end{equation}
Note $\widetilde{\mathcal{V}}_{0}=L_{\widetilde{w}}V_{T^{4}}\Omega_{3}$.
Since the fields are independent of $\widetilde{w}$ and $\eta$ is
closed, 
\begin{equation}
\partial_{I}\left(\sqrt{\widetilde{h}}q_{\mathrm{D2}}^{I}\right)=0.\label{eq:D2-effective-conservation}
\end{equation}
For the spatial metric in Eq. (\ref{eq:D0D4F1-metric}), we have

\begin{equation}
\sqrt{\widetilde{h}}=\sqrt{\widetilde{g}_{rr}}Y_{\mathrm{D2}}\left(r\right)\sqrt{\bar{\gamma}},\qquad Y_{\mathrm{D2}}\left(r\right)=\sqrt{\widetilde{g}_{\widetilde{w}\widetilde{w}}}\left(\sqrt{\widetilde{g}_{S^{3}}}\right)^{3}\sqrt{\widetilde{g}_{T^{4}}}=r^{3}H_{1}^{2}H_{P}^{-1/2}.\label{eq:D2-Y}
\end{equation}
For comparison, the Type IIB D1--D5--P area density is 
\begin{equation}
Y_{\mathrm{D1}}\left(r\right)=r^{3}H_{1}^{3/2}H_{5}^{-1/2}H_{P}^{1/2}.\label{eq:D1-Y}
\end{equation}
Because the original metric (\ref{eq:RR-components-before-T}) has
$g_{ww}=H_{P}/\sqrt{H_{1}H_{5}}$, 
\begin{equation}
Y_{\mathrm{D2}}=\frac{Y_{\mathrm{D1}}}{g_{ww}},\label{eq:D2-Y-transform}
\end{equation}
again in agreement with the general Buscher relation. We now define
the bit-thread flow
\begin{equation}
v_{\mathrm{D2}}^{I}=C_{{\rm geom}}^{\mathrm{D2}}q_{\mathrm{D2}}^{I}.\label{eq:D2-flow-definition}
\end{equation}
Its radial component and norm are 

\begin{equation}
v_{\mathrm{D2}}^{r}=\frac{C_{{\rm geom}}^{\mathrm{D2}}}{\sqrt{\widetilde{g}_{rr}}Y_{\mathrm{D2}}\left(r\right)}\frac{\mathcal{N}}{\widetilde{\mathcal{V}}_{0}},\qquad\left|v_{\mathrm{D2}}\left(r\right)\right|=\sqrt{\widetilde{h}_{IJ}v_{\mathrm{D2}}^{I}v_{\mathrm{D2}}^{J}}=\frac{C_{{\rm geom}}^{\mathrm{D2}}}{Y_{\mathrm{D2}}\left(r\right)}\frac{\mathcal{N}}{\widetilde{\mathcal{V}}_{0}}.\label{eq:D2-flow-and-norm}
\end{equation}
Using Eq. (\ref{eq:D0D4F1-dilaton}), 

\begin{align}
e^{-2\widetilde{\phi}_{\mathrm{D2}}}Y_{\mathrm{D2}} & =H_{1}^{-3/2}H_{5}^{1/2}H_{P}\left(r^{3}H_{1}^{2}H_{P}^{-1/2}\right)\nonumber \\
 & =r^{3}\sqrt{H_{1}H_{5}H_{P}}=\sqrt{\left(r^{2}+Q_{1}\right)\left(r^{2}+Q_{5}\right)\left(r^{2}+Q_{P}\right)}.\label{eq:D2-weighted-area}
\end{align}
Saturation at $r=0$ fixes 
\begin{equation}
C_{{\rm geom}}^{\mathrm{D2}}=\frac{\widetilde{\mathcal{V}}_{0}}{4\widetilde{G}_{N}^{\left(10\right)}\mathcal{N}}\sqrt{Q_{1}Q_{5}Q_{P}}.\label{eq:D2-Cgeom}
\end{equation}
The pointwise bound is 
\begin{equation}
\frac{\left|v_{\mathrm{D2}}\left(r\right)\right|}{e^{-2\widetilde{\phi}_{\mathrm{D2}}}/\left(4\widetilde{G}_{N}^{\left(10\right)}\right)}=\left[\frac{Q_{1}Q_{5}Q_{P}}{\left(r^{2}+Q_{1}\right)\left(r^{2}+Q_{5}\right)\left(r^{2}+Q_{P}\right)}\right]^{1/2}\leq1.\label{eq:D2-pointwise-bound}
\end{equation}
The reduced D2 current therefore produces a valid radial entropy flow
throughout the exterior. It saturates the capacity at the horizon
and is maximal with respect to the horizon cut. Its flux is 
\begin{align}
\Phi_{\mathrm{D2}} & =\int_{\widetilde{\Sigma}_{8}}v_{\mathrm{D2}}^{r}\widetilde{n}_{r}d\widetilde{A}=C_{{\rm geom}}^{\mathrm{D2}}\mathcal{N}\nonumber \\
 & =\frac{L_{\widetilde{w}}V_{T^{4}}\Omega_{3}}{4\widetilde{G}_{N}^{\left(10\right)}}\sqrt{Q_{1}Q_{5}Q_{P}}=S_{\mathrm{BH}}^{\mathrm{D0-D4-F1}}.\label{eq:D2-entropy-flux}
\end{align}
Thus, with the geometric normalization fixed by horizon saturation,
the maximal flux of the topologically reduced D2 current equals the
D0-D4-F1 black hole entropy. 

Moreover, recall the Type IIB D1--D5--P charge parameters \cite{Peet:2000hn}:
\begin{equation}
Q_{1}=\frac{g_{s}\alpha^{\prime3}}{V_{4}}n_{1},\qquad Q_{5}=g_{s}\alpha^{\prime}n_{5},\qquad Q_{P}=\frac{g_{s}^{2}\alpha^{\prime4}}{R_{w}^{2}V_{4}}n_{P}.\label{eq:D1D5P-charge-dictionary}
\end{equation}
After T-duality, the Type IIA charges are 
\begin{equation}
\widetilde{Q}_{0}=\frac{\widetilde{g}_{s}\alpha^{\prime7\text{/2}}}{R_{\widetilde{w}}V_{4}}\widetilde{n}_{0},\qquad\widetilde{Q}_{4}=\frac{\widetilde{g}_{s}\alpha^{\prime3/2}}{R_{\widetilde{w}}}\widetilde{n}_{4},\qquad\widetilde{Q}_{\mathrm{F1}}=\frac{\widetilde{g}_{s}^{2}\alpha^{\prime3}}{V_{4}}\widetilde{n}_{\mathrm{F1}}.\label{eq:D0D4F1-charge-dictionary}
\end{equation}
Using Eq. (\ref{eq:D0D4F1-charge-dictionary}), 
\begin{equation}
\sqrt{\widetilde{Q}_{0}\widetilde{Q}_{4}\widetilde{Q}_{\mathrm{F1}}}=\frac{\widetilde{g}_{s}^{2}\alpha^{\prime4}}{R_{\widetilde{w}}V_{4}}\sqrt{\widetilde{n}_{0}\widetilde{n}_{4}\widetilde{n}_{\mathrm{F1}}}.\label{eq:app-RR-charge-product}
\end{equation}
Moreover, 
\begin{equation}
\widetilde{\mathcal{V}}_{0}=\left(2\pi R_{\widetilde{w}}\right)\left(2\pi\right)^{4}V_{4}\left(2\pi^{2}\right),\qquad4\widetilde{G}_{N}^{\left(10\right)}=32\pi^{6}\widetilde{g}_{s}^{2}\alpha^{\prime4}.\label{eq:app-volume-G}
\end{equation}
Therefore, 
\begin{align}
\frac{\widetilde{\mathcal{V}}_{0}}{4\widetilde{G}_{N}^{\left(10\right)}}\sqrt{\widetilde{Q}_{0}\widetilde{Q}_{4}\widetilde{Q}_{\mathrm{F1}}} & =\frac{\left(2\pi R_{\widetilde{w}}\right)\left(2\pi\right)^{4}V_{4}\left(2\pi^{2}\right)}{32\pi^{6}\widetilde{g}_{s}^{2}\alpha^{\prime4}}\frac{\widetilde{g}_{s}^{2}\alpha^{\prime4}}{R_{\widetilde{w}}V_{4}}\sqrt{\widetilde{n}_{0}\widetilde{n}_{4}\widetilde{n}_{\mathrm{F1}}}\nonumber \\
 & =2\pi\sqrt{\widetilde{n}_{0}\widetilde{n}_{4}\widetilde{n}_{\mathrm{F1}}}.\label{eq:app-RR-integer-entropy}
\end{align}
The NS--NS calculation is identical after replacing $(\widetilde{n}_{0},\widetilde{n}_{4},\widetilde{n}_{\mathrm{F1}})$
by $(\widetilde{n}_{\mathrm{F1}},\widetilde{n}_{\mathrm{NS5}},\widetilde{n}_{\mathrm{P}})$.
Thus, the factors of $R_{\widetilde{w}}$ in the D0 and D4 charges
reflect their smearing along the dual circle. With 
\begin{equation}
\left(\widetilde{n}_{0},\widetilde{n}_{4},\widetilde{n}_{\mathrm{F1}}\right)=\left(n_{1},n_{5},n_{P}\right),\label{eq:D0D4F1-integer-map}
\end{equation}
and the modulus transformation (\ref{eq:T-dual-moduli}),

\begin{equation}
R_{\widetilde{w}}=\frac{\alpha^{\prime}}{R_{w}},\qquad\widetilde{g}_{s}=g_{s}\frac{\sqrt{\alpha^{\prime}}}{R_{w}},
\end{equation}
we have

\begin{equation}
\widetilde{Q}_{0}=Q_{1},\qquad\widetilde{Q}_{4}=Q_{5},\qquad\widetilde{Q}_{\mathrm{F1}}=Q_{P}.\label{eq:D0D4F1-Q-map}
\end{equation}
Eq. (\ref{eq:D2-entropy-flux}) becomes 
\begin{equation}
S_{\mathrm{BH}}^{\mathrm{D0-D4-F1}}=2\pi\sqrt{\widetilde{n}_{0}\widetilde{n}_{4}\widetilde{n}_{\mathrm{F1}}}=2\pi\sqrt{n_{1}n_{5}n_{P}}.\label{eq:D0D4F1-integer-entropy}
\end{equation}
This agrees with the standard D1--D5--P entropy \cite{Strominger:1996sh,Callan:1996dv}.
It also shows that the entropy is unchanged under the T-duality map.

\section{Local and global T-duality invariants}

\label{sec:invariants}

We now compare the four radial flows as follows. The source carriers
and their T-dual images are summarized in Table \ref{tab:duality-summary}.

\begin{table}[h]
\centering %
\begin{tabular}{@{}llll@{}}
\toprule 
Original background & Original radial carrier & T-dual background & T-dual radial carrier\tabularnewline
\midrule 
IIB F1--NS5--P & F1, $j^{0r}$ & IIA P--NS5--F1 & F1, $\widetilde{j}^{0r}$\tabularnewline
IIB D1--D5--P & D1, $j^{0r}$ & IIA D0--D4--F1 & wrapped D2, $\widetilde{j}^{0r\widetilde{w}}$\tabularnewline
\bottomrule
\end{tabular}\caption{The background and auxiliary radial-source maps under T-duality along
the common circle. In the second row, the bit-thread vector is obtained
from $j^{0r\widetilde{w}}$ by contraction with $\eta=d\widetilde{w}/L_{\widetilde{w}}$.}
\label{tab:duality-summary}
\end{table}
In the D1--D2 chain, the wrapped D2 current is first projected onto
the spatial bivector $Q_{\mathrm{D2}}^{r\widetilde{w}}$ and then
contracted with the normalized one-form 
\begin{equation}
\eta=\frac{d\widetilde{w}}{L_{\widetilde{w}}},
\end{equation}
to obtain the effective radial vector.

Although the four flows describe the same physical entropy flux, their
local expressions differ. The string-frame transverse area densities
are 
\begin{align}
Y_{\mathrm{F1}}^{\mathrm{IIB}} & =r^{3}H_{1}^{-1/2}H_{5}^{3/2}H_{P}^{1/2}, & Y_{\mathrm{F1}}^{\mathrm{IIA}} & =r^{3}H_{P}^{-1/2}H_{5}^{3/2}H_{1}^{1/2},\\
Y_{\mathrm{D1}}^{\mathrm{IIB}} & =r^{3}H_{1}^{3/2}H_{5}^{-1/2}H_{P}^{1/2}, & Y_{\mathrm{D2}}^{\mathrm{IIA}} & =r^{3}H_{1}^{2}H_{P}^{-1/2}.
\end{align}
These quantities differ because the proper sizes of $S^{1}$, $S^{3}$,
and $T^{4}$ depend on the duality frame. The corresponding dilaton
factors are 
\begin{align}
e^{-2\phi_{\mathrm{F1}}^{\mathrm{IIB}}} & =\frac{H_{1}}{H_{5}}, & e^{-2\phi_{\mathrm{F1}}^{\mathrm{IIA}}} & =\frac{H_{P}}{H_{5}},\\
e^{-2\phi_{\mathrm{D1}}^{\mathrm{IIB}}} & =\frac{H_{5}}{H_{1}}, & e^{-2\phi_{\mathrm{D2}}^{\mathrm{IIA}}} & =H_{1}^{-3/2}H_{5}^{1/2}H_{P}.
\end{align}
Nevertheless, the dilaton-weighted area density is identical in all
four frames: 

\begin{equation}
e^{-2\phi_{\mathrm{F1}}^{\mathrm{IIB}}}Y_{\mathrm{F1}}^{\mathrm{IIB}}=e^{-2\phi_{\mathrm{F1}}^{\mathrm{IIA}}}Y_{\mathrm{F1}}^{\mathrm{IIA}}=e^{-2\phi_{\mathrm{D1}}^{\mathrm{IIB}}}Y_{\mathrm{D1}}^{\mathrm{IIB}}=e^{-2\phi_{\mathrm{D2}}^{\mathrm{IIA}}}Y_{\mathrm{D2}}^{\mathrm{IIA}}=r^{3}\sqrt{H_{1}H_{5}H_{P}}.\label{eq:universal-weighted-area-density}
\end{equation}
The equality between the Type IIB and Type IIA F1 expressions follows
from NS--NS T-duality. Similarly, the equality between the Type IIB
D1 and Type IIA D2 expressions follows from the T-duality from D1-brane
to D2-brane. The equality between the Type IIB F1 and D1 expressions
is the S-duality invariant found in our previous analysis. Together,
these relations show that the weighted area density is the same in
all four duality frames.

Moreover, the geometric conversion coefficients satisfy 
\begin{equation}
\widetilde{C}_{{\rm geom}}=\frac{\widetilde{\mathcal{V}}_{0}}{4\widetilde{G}_{N}^{\left(10\right)}\mathcal{N}}\sqrt{Q_{1}Q_{5}Q_{P}}=\frac{\mathcal{V}_{0}}{4G_{N}^{\left(10\right)}\mathcal{N}}\sqrt{Q_{1}Q_{5}Q_{P}}=C_{{\rm geom}},\label{eq:Cgeom-T-invariant}
\end{equation}

\noindent where Eq. (\ref{eq:global-invariant-intro}) has been used.
This equality assumes the one-to-one map of the probe number $\mathcal{N}$
and the same unit-charge convention in both frames; for the D2 carrier
it also uses $\int_{S_{\widetilde{w}}^{1}}\eta=1$. Recall the physical
entropy flux:
\begin{equation}
\Phi=C_{\mathrm{geom}}\mathcal{N}=\frac{\mathcal{V}_{0}}{4G_{N}^{\left(10\right)}}\sqrt{Q_{1}Q_{5}Q_{P}},\label{eq:flux-before-T}
\end{equation}
whereas in the T-dual frame, 
\begin{equation}
\widetilde{\Phi}=\widetilde{C}_{\mathrm{geom}}\mathcal{N}=\frac{\widetilde{\mathcal{V}}_{0}}{4\widetilde{G}_{N}^{\left(10\right)}}\sqrt{Q_{1}Q_{5}Q_{P}}.\label{eq:flux-after-T}
\end{equation}
Therefore, 
\begin{equation}
\Phi_{\mathrm{F1}}^{\mathrm{IIB}}=\Phi_{\mathrm{F1}}^{\mathrm{IIA}}=\Phi_{\mathrm{D1}}^{\mathrm{IIB}}=\Phi_{\mathrm{D2}}^{\mathrm{IIA}}=S_{\mathrm{BH}}.\label{eq:four-frame-flux-invariant}
\end{equation}
The carrier can be an NS-NS string current, an R-R string current,
or a wrapped R-R brane current. No individual current component or
vector norm is invariant. The local invariant is the dimensionless
ratio of the flow norm to the corresponding string-frame capacity,
while the global invariant is the maximal physical flux obtained after
reducing the source current to the appropriate one-dimensional conserved
flow.

Finally, the D1--D5 CFT fixes the microscopic entropy:
\begin{equation}
S_{{\rm CFT}}=2\pi\sqrt{n_{1}n_{5}n_{P}}.\label{eq:CFT-entropy-short}
\end{equation}
The \emph{boundary-matching conjecture} of Ref. \cite{Wu:2026qha}
assigns this entropy to the integrated radial D1 flow of one complete
boundary factor. For a fixed coarse-graining into $\mathcal{N}$ source
tubes, it gives 
\begin{equation}
C_{{\rm CFT}}=\frac{S_{{\rm CFT}}}{\mathcal{N}}.\label{eq:CFT-C-short}
\end{equation}
Conditional on this conjecture and the fixed tube normalization, the
coefficient is carried through T-duality without change:
\begin{equation}
\widetilde{C}_{{\rm CFT}}=C_{{\rm CFT}}=C_{{\rm geom}}=\widetilde{C}_{{\rm geom}}.\label{eq:CFT-C-T-dual}
\end{equation}
Consequently, if the \emph{boundary-matching conjecture} is imposed
in the D1-D5-P frame, T-duality predicts the capacity of the wrapped-D2
tubes in the D0-D4-F1 frame.

It is worth noting that the T-duality along the BTZ circle maps the
near-horizon geometry to the three-dimensional charged black string,
whose asymptotic structure is not the standard AdS$_{3}$ boundary
used in the D1-D5 CFT description. What is established here is the
transport of the already fixed conserved flux through an exact bulk
duality. A direct boundary formulation of the charged black string
flow would require a separate analysis.

\section{Discussion and conclusion}

\label{sec:conclusion}

In this paper, we studied T-duality of the string charge density/bit
threads correspondence in two three-charge systems. In the first duality
route, Type IIB F1--NS5--P is mapped to a Type IIA P--NS5--F1
background. The Buscher rules exchange the roles of the fundamental-string
and momentum harmonic functions. A radial F1 source remains an F1
source, and the maximal flux of its projected charge density reproduces
the entropy of the T-dual black hole.

In the second route, Type IIB D1--D5--P is mapped to Type IIA D0--D4--F1.
Here T-duality changes the dimension of the auxiliary radial carrier.
A D1 transverse to the duality circle becomes a D2 wrapped on the
dual circle. Its projected charge density is a spatial bivector rather
than a vector. The normalized one-form $\eta=d\widetilde{w}/L_{\widetilde{w}}$
provides a canonical reduction, 
\begin{equation}
q_{{\rm D2}}^{I}=Q_{\mathrm{D2}}^{IJ}\eta_{J}.\label{eq:conclusion-effective-flow}
\end{equation}
The resulting vector is divergenceless, has total source flux $\mathcal{N}$,
and defines a valid bit-thread flow. It is the reduced flow of the
wrapped-D2 current. Its maximal entropy flux $\Phi_{\mathrm{D2}}^{\mathrm{IIA}}$
equals the D0--D4--F1 Bekenstein--Hawking entropy.

This result clarifies which part of the proposed correspondence is
fundamental. The rank and local form of the source current are not
invariant. What is preserved is the effective one-dimensional charge
density flow obtained after reducing every wrapped worldvolume direction
with a topologically normalized form. More generally, for a D$\left(k+1\right)$-brane
wrapped on a product of $k$ circles with closed normalized one-forms
$\eta^{\left(a\right)}$, the natural candidate radial vector is 
\begin{equation}
q_{{\rm D\left(k+1\right)}}^{I}=Q^{IJ_{1}\cdots J_{k}}\eta_{J_{1}}^{\left(1\right)}\cdots\eta_{J_{k}}^{\left(k\right)}.\label{eq:general-wrapped-brane-reduction}
\end{equation}
For a general compact cycle, the analogous construction should instead
contract the brane current with a closed normalized $k$-form. Establishing
conservation for general cycles and nontrivial fibrations would extend
the correspondence from string currents to a broader class of T-dual
wrapped branes.

The calculations also identify two complementary invariants. Locally,
\begin{equation}
e^{-2\phi}Y=r^{3}\sqrt{H_{1}H_{5}H_{P}},\label{eq:conclusion-local-invariant}
\end{equation}
is common to the four frames. It determines both the horizon bottleneck
and the complete radial profile relative to the norm bound. Globally,
\begin{equation}
\frac{\widetilde{\mathcal{V}}_{0}}{\widetilde{G}_{N}^{\left(10\right)}}=\frac{\mathcal{V}_{0}}{G_{N}^{\left(10\right)}},\label{eq:conclusion-global-invariant}
\end{equation}
ensures that the integrated maximal flux is unchanged. These identities
explain how T-duality can change the string-frame geometry and the
flow norm while preserving the entropy.

It is worth noting that the analysis assumes an Abelian isometry and
uses source ensembles smeared along the dualized circle. This is required
by the zero-mode Buscher rules. Moreover, the radial sources are probes
and do not change the three-charge background at leading order. On
the other hand, the calculation uses the two-derivative supergravity
action. Extending the correspondence beyond this approximation requires
the $\alpha^{\prime}$-corrected T-duality rules and the corresponding
higher-curvature bit-thread capacity. For example, in the NS-NS sector
of Type II theory at order $\alpha^{\prime3}$, the $O\left(d,d\right)$
structure of the dimensionally reduced action can be recovered through
field redefinitions, but the required field redefinitions cannot,
in general, be uplifted to local ten-dimensional ones \cite{Hsia:2024kpi}.
This obstruction does not affect the leading-order analysis performed
here, but it constrains how a higher-derivative duality-covariant
extension of the correspondence can be formulated. Finally, we have
studied only a single circle duality. It would be interesting to organize
the compactified metric, Kalb-Ramond field, compactification measure,
and reduced brane currents into $O\left(d,d\right)$ multiplets directly
in the dimensionally reduced theory.

Under these assumptions, the result provides a nontrivial leading-order
T-duality test of the string charge density/bit threads correspondence.
With the geometric normalization fixed by the same horizon-saturation
prescription in each frame, the Type IIA F1 and wrapped-D2 currents
have maximal fluxes equal to the corresponding black hole entropies.
This supports an information-flow picture in which different string
and brane currents are duality-frame-dependent carriers of the same
conserved maximal entropy flux.

\vspace{5mm}

\noindent {\bf Acknowledgements} 
HW is supported by NSFC Grant No.12105191. SY is supported by NSFC Grant No.12105031 and No.12547101.


\begin{thebibliography}{99}


\bibitem{Ryu:2006bv} S.~Ryu and T.~Takayanagi,   ``Holographic derivation of entanglement entropy from AdS/CFT,''   Phys.\ Rev.\ Lett.\  {\bf 96}, 181602 (2006)   doi:10.1103/PhysRevLett.96.181602   [hep-th/0603001].   

\bibitem{Ryu:2006ef}    S.~Ryu and T.~Takayanagi,   ``Aspects of Holographic Entanglement Entropy,''   JHEP {\bf 0608}, 045 (2006)   doi:10.1088/1126-6708/2006/08/045   [hep-th/0605073].   

 



\bibitem{Freedman:2016zud} M.~Freedman and M.~Headrick, ``Bit threads and holographic entanglement,'' Commun. Math. Phys. \textbf{352}, no.1, 407-438 (2017) doi:10.1007/s00220-016-2796-3 [arXiv:1604.00354 [hep-th]]. 

\bibitem{Headrick:2017ucz} M.~Headrick and V.~E.~Hubeny, ``Riemannian and Lorentzian flow-cut theorems,'' Class. Quant. Grav. \textbf{35}, no.10, 105012 (2018) doi:10.1088/1361-6382/aab83c [arXiv:1710.09516 [hep-th]]. 




\bibitem{Agon:2018lwq} C.~A.~Ag{\'o}n, J.~De Boer and J.~F.~Pedraza, ``Geometric Aspects of Holographic Bit Threads,'' JHEP \textbf{05}, 075 (2019) doi:10.1007/JHEP05(2019)075 [arXiv:1811.08879 [hep-th]]. 

\bibitem{Caggioli:2024uza} S.~Caggioli, F.~Gentile, D.~Seminara and E.~Tonni, ``Holographic thermal entropy from geodesic bit threads,'' JHEP \textbf{07}, 088 (2024) doi:10.1007/JHEP07(2024)088 [arXiv:2403.03930 [hep-th]]. 

\bibitem{Lin:2026ehl} Y.~Y.~Lin and S.~Cheng, ``Holographic Bit Threads from String-Diagrammatic Quantum Information Flow,'' [arXiv:2608.19428 [hep-th]]. 

\bibitem{Caceres:2025ypk} E.~C{\'a}ceres, R.~Carrasco and J.~F.~Pedraza, ``Lorentzian threads and nonlocal computation in holography,'' Phys. Rev. D \textbf{113}, no.10, 106024 (2026) doi:10.1103/bk1l-3ffb [arXiv:2512.07963 [hep-th]]. 






\bibitem{Wu:2025qwc} H.~Wu and S.~Ying, ``Toward a worldsheet theory of entanglement entropy,'' Phys. Rev. D \textbf{113}, no.6, 066006 (2026) doi:10.1103/cgg9-p6xy [arXiv:2511.16586 [hep-th]]. 


\bibitem{Wu:2026qha} H.~Wu and S.~Ying, ``String charge density/bit threads correspondence and its S-duality in type IIB supergravity,'' [arXiv:2609.15929 [hep-th]]. 





\bibitem{Bergshoeff:1996tu} E.~Bergshoeff and P.~K.~Townsend, ``Super D-branes,'' Nucl. Phys. B \textbf{490}, 145-162 (1997) doi:10.1016/S0550-3213(97)00072-2 [arXiv:hep-th/9611173 [hep-th]]. 

\bibitem{Hassan:1999bv} S.~F.~Hassan, ``T duality, space-time spinors and RR fields in curved backgrounds,'' Nucl. Phys. B \textbf{568}, 145-161 (2000) doi:10.1016/S0550-3213(99)00684-7 [arXiv:hep-th/9907152 [hep-th]]. 





\bibitem{He:2014gva} S.~He, T.~Numasawa, T.~Takayanagi and K.~Watanabe, ``Notes on Entanglement Entropy in String Theory,'' JHEP \textbf{05}, 106 (2015) doi:10.1007/JHEP05(2015)106 [arXiv:1412.5606 [hep-th]]. 

\bibitem{Ahmadain:2022tew} 
A.~Ahmadain and A.~C.~Wall, ``Off-shell strings I: S-matrix and action,'' SciPost Phys. \textbf{17}, no.1, 005 (2024) doi:10.21468/SciPostPhys.17.1.005 [arXiv:2211.08607 [hep-th]]. 


\bibitem{Ahmadain:2022eso} 
A.~Ahmadain and A.~C.~Wall, ``Off-shell strings II: Black hole entropy,'' SciPost Phys. \textbf{17}, no.1, 006 (2024) doi:10.21468/SciPostPhys.17.1.006 [arXiv:2211.16448 [hep-th]]. 


\bibitem{Brustein:2022wiq} R.~Brustein and Y.~Zigdon, ``On the entropy of strings and branes,'' JHEP \textbf{10}, 112 (2022) doi:10.1007/JHEP10(2022)112 [arXiv:2208.07372 [hep-th]]. 


\bibitem{Halder:2023adw} I.~Halder and D.~L.~Jafferis, ``Thermal Bekenstein-Hawking entropy from the worldsheet,'' JHEP \textbf{05}, 136 (2024) doi:10.1007/JHEP05(2024)136 [arXiv:2310.02313 [hep-th]]. 

\bibitem{Ahmadain:2024hdp} A.~Ahmadain, A.~Frenkel and A.~C.~Wall, ``A Background-Independent Closed String Action at Tree Level,'' [arXiv:2410.11938 [hep-th]]. 



\bibitem{Mori:2025qoh} S.~Mori, T.~Sakai and M.~Shigemori, ``Black Hole Entropy from String Entanglement,'' [arXiv:2509.21796 [hep-th]]. 

\bibitem{Ahmadain:2025pox} A.~Ahmadain and M.~Yang, ``Strings at the Tip of the Cone and Black Hole Entropy From the Worldsheet: Part I,'' [arXiv:2512.00637 [hep-th]]. 

\bibitem{Jorstad:2026jlg} E.~J{\o}rstad, R.~C.~Myers and S.~Pasterski, ``Flat Space Entanglement: A Coulomb Branch Perspective,'' [arXiv:2606.13889 [hep-th]]. 











\bibitem{Tseytlin:1996bh} A.~A.~Tseytlin, ``Harmonic superpositions of M-branes,'' Nucl. Phys. B \textbf{475}, 149-163 (1996) doi:10.1016/0550-3213(96)00328-8 [arXiv:hep-th/9604035 [hep-th]]. 

\bibitem{Mathur:2005zp} S.~D.~Mathur, ``The Fuzzball proposal for black holes: An Elementary review,'' Fortsch. Phys. \textbf{53}, 793-827 (2005) doi:10.1002/prop.200410203 [arXiv:hep-th/0502050 [hep-th]]. 




\bibitem{Buscher:1987sk} T.~H.~Buscher, ``A Symmetry of the String Background Field Equations,'' Phys. Lett. B \textbf{194}, 59-62 (1987) doi:10.1016/0370-2693(87)90769-6 

\bibitem{Buscher:1987qj} T.~H.~Buscher, ``Path Integral Derivation of Quantum Duality in Nonlinear Sigma Models,'' Phys. Lett. B \textbf{201}, 466-472 (1988) doi:10.1016/0370-2693(88)90602-8 

\bibitem{Alvarez:1994dn} E.~Alvarez, L.~Alvarez-Gaume and Y.~Lozano, ``An Introduction to T duality in string theory,'' Nucl. Phys. B Proc. Suppl. \textbf{41}, 1-20 (1995) doi:10.1016/0920-5632(95)00429-D [arXiv:hep-th/9410237 [hep-th]]. 





\bibitem{Bena:2022sge} I.~Bena, N.~Ceplak, S.~Hampton, Y.~Li, D.~Toulikas and N.~P.~Warner, ``Resolving black-hole microstructure with new momentum carriers,'' JHEP \textbf{10}, 033 (2022) doi:10.1007/JHEP10(2022)033 [arXiv:2202.08844 [hep-th]]. 


\bibitem{Maldacena:1996ix} J.~M.~Maldacena and A.~Strominger, ``Black hole grey body factors and d-brane spectroscopy,'' Phys. Rev. D \textbf{55}, 861-870 (1997) doi:10.1103/PhysRevD.55.861 [arXiv:hep-th/9609026 [hep-th]]. 



\bibitem{Myers:1999ps} R.~C.~Myers, ``Dielectric branes,'' JHEP \textbf{12}, 022 (1999) doi:10.1088/1126-6708/1999/12/022 [arXiv:hep-th/9910053 [hep-th]]. 





\bibitem{Bena:2008dw} I.~Bena, N.~Bobev, C.~Ruef and N.~P.~Warner, ``Supertubes in Bubbling Backgrounds: Born-Infeld Meets Supergravity,'' JHEP \textbf{07}, 106 (2009) doi:10.1088/1126-6708/2009/07/106 [arXiv:0812.2942 [hep-th]]. 






\bibitem{Frey:2019fqz} A.~R.~Frey, ``Dirac branes for Dirichlet branes: Supergravity actions,'' Phys. Rev. D \textbf{102}, no.4, 046017 (2020) doi:10.1103/PhysRevD.102.046017 [arXiv:1907.12755 [hep-th]]. 


\bibitem{Blau:2002} M.~Blau, ``Supergravity Solitons,'' Lectures presented at the Introductory School on String Theory, ICTP, Trieste, 3--14 June 2002, \url{http://www.ictp.trieste.it/~mblau/}.





\bibitem{Peet:2000hn} A.~W.~Peet, ``TASI lectures on black holes in string theory,'' doi:10.1142/9789812799630{\_}0003 [arXiv:hep-th/0008241 [hep-th]]. 







\bibitem{Strominger:1996sh} A.~Strominger and C.~Vafa, ``Microscopic origin of the Bekenstein-Hawking entropy,'' Phys. Lett. B \textbf{379}, 99-104 (1996) doi:10.1016/0370-2693(96)00345-0 [arXiv:hep-th/9601029 [hep-th]]. 

\bibitem{Callan:1996dv} C.~G.~Callan and J.~M.~Maldacena, ``D-brane approach to black hole quantum mechanics,'' Nucl. Phys. B \textbf{472}, 591-610 (1996) doi:10.1016/0550-3213(96)00225-8 [arXiv:hep-th/9602043 [hep-th]]. 



\bibitem{Hsia:2024kpi} S.~W.~Hsia, A.~R.~Kamal and L.~Wulff, ``No manifest T duality at order {\ensuremath{\alpha}}'3,'' Phys. Rev. D \textbf{111}, no.6, L061904 (2025) doi:10.1103/PhysRevD.111.L061904 [arXiv:2411.15302 [hep-th]]. 

\end{thebibliography}
\end{document}